\documentclass[english,10pt,journal,twocolumn]{IEEEtran}
\usepackage[utf8]{inputenc}
\usepackage{amsthm}
\usepackage{amsfonts}
\usepackage{amssymb}
\usepackage{graphicx}
\usepackage{multicol}
\usepackage{epstopdf}
\renewcommand{\vec}[1]{\mathbf{#1}}
\usepackage[cmex10]{amsmath}
\usepackage[justification=centering]{caption}
\usepackage{enumerate} 
\usepackage{cite}
\usepackage[dvipsnames]{xcolor}
\newcommand{\vast}{\bBigg@{4}}
\newcommand{\Vast}{\bBigg@{5}}
\usepackage[list=true]{subcaption}

\begin{document}

\title{Analysis of Optimal Combining in Rician Fading with Co-channel Interference}

\author{Muralikrishnan Srinivasan, Sheetal Kalyani\\
\thanks{The authors are with the Department of Electrical Engineering, Indian Institute of Technology Madras,  Chennai, India 600036 (email:\{ee14d206,skalyani\}@ee.iitm.ac.in). }
\thanks{Copyright (c) 2019 IEEE. Personal use is permitted. For any other purposes, permission must be obtained from the IEEE by emailing pubs-permissions@ieee.org. This is the author's version of an article that has been published in this journal. Changes were made to this version by the publisher prior to publication. The final version of record is available at http://dx.doi.org/10.1109/TVT.2019.2898235
} 
}
 \maketitle

 \begin{abstract}
Approximate Symbol error rate (SER), outage probability and rate expressions are derived for receive diversity system employing optimum combining when both the desired and the interfering signals are subjected to Rician fading, for the cases of a) equal power uncorrelated interferers b) unequal power interferers c) interferer correlation. The derived expressions are applicable for an arbitrary number of receive antennas and interferers and for any quadrature amplitude modulation (QAM) constellation. Furthermore, we derive a simple closed form expression for SER in the interference-limited regime, for the special case of Rayleigh faded interferers.  A close match is observed between the SER, outage probability and rate results obtained through the derived analytical expressions and the ones obtained from Monte-Carlo simulations. 
\end{abstract}


\IEEEpeerreviewmaketitle

\section{Introduction}
\IEEEPARstart{A}{mong} the various diversity combining schemes, optimum combining (OC) proposed in \cite{winters} maximizes the signal to interference plus noise ratio (SINR). Performance of OC receivers has been extensively studied for various cases when both the desired and interfering signals are subjected to Rayleigh fading \cite{chiric, ali, rahim, chung, shah, kwak, chi1, mallik, sri, yue, zhang, afana}. Many practical scenarios exist such as indoor propagation, micro-cellular channels, satellite channels, inter-vehicular communications, etc, where both the desired and interfering signals may have line-of-sight (LoS) paths. Symbol error rate (SER) expressions for  OC have been derived, when either the desired signals or the interfering signals undergo Rician fading, while the other undergoes Rayleigh fading \cite{chiric}. 
\par Rician fading has found applications even in recent times in the study of the performance of distributed multiple input multiple output (MIMO) with zero forcing (ZF) receivers over correlated Rician fading channels \cite{li_mimo}, in deriving expressions for achievable rates of MIMO relay systems with ZF processing over Rician fading systems \cite{wang_mimo} and in the study of the performance of co-operative relaying systems with non-orthogonal multiple access \cite{jiao_noma}. Existing 4G and emerging 5G systems are both interference-limited. Hence, receiver techniques like OC and MRC will play a key role in the performance analysis of these systems \cite{ali_lcr}. However, SER and/or outage probability (OP) expressions when OC is employed and when both the user and interferers undergo Rician fading have not been studied.  Further, characterization of OC receivers, which takes into account practical scenarios such as unequal interference power and correlation among interferers is not present in open literature. Hence, we address the gap in the literature with the following contributions through this paper:
\begin{itemize}
    \item We derive exact expressions for the Laplace transform of SINR at the output of OC considering Rician faded users and a) mixture of Rayleigh and Rician faded interferers, b) only Rician faded interferers and c) only Rayleigh faded interferers. A simple approximation, which avoids determinant evaluation, is also derived for Rayleigh faded interferers  in an interference dominated scenario.
    \item We also derive exact Laplace transform expressions of SINR for unequal power interferers and correlated interferers, which occur due to correlated channel fading, shadowing and from spatial distribution of transmitters \cite{zhong_correlation, kumar_impact, kumar_ffr}. An extension to $\kappa-\mu$ faded users is also proposed.
    \item Using these Laplace transform expressions, we derive approximate SER expressions that are functions of a double infinite series, which are truncated to finite series with arbitrarily small truncation error. The series terms are functions of Tricomi hypergeometric functions, which has been used extensively in analyzing throughput and the rate of wireless systems over various fading channels \cite{li_miso, hanzo, you, zhang1}. 
    \item We also derive an expression for the moments of the SINR $\eta$. The first two moments are then matched with those of a beta-prime random variable to obtain approximate outage probability and rate expressions. Inferences on the impact of the fading parameters are analytically studied by using stochastic ordering tools on the outage and SER expressions. All our results are compared with corresponding Monte-Carlo simulations and a close match is observed. We also give an application of OC in vehicular technology networks.
\end{itemize}
The notations used in the paper are: $\mathcal{CN}(.,.)$ denotes complex normal random variable, $(.)^H$ denotes transpose of a matrix, $\mathbb{E}(.)$ denotes expectation, $\,_1F_1(.)$ denotes confluent hypergeometric function, $|.|$ denotes determinant of a matrix, tr$(.)$ denotes trace, $\mathcal{CW}(.)$ denotes a complex Wishart random matrix, $U(.)$ denotes Tricomi hypergeometric function, $\otimes$ denotes Kronecker product, $\Psi(.)$ is the digamma function.
\section{ System Model}
Let $N_R$ denote the number of receive antennas, $N_I$ denote the number of interferers,
$\mathbf{c}$ denote the $N_R \times 1$ channel from the transmitter to the user, $\mathbf{c}_i$ denote the  $N_R \times 1$ channel from the $i^{th}$ interferer to the user, $x$ denotes the desired user symbol belonging to unit-energy quadrature amplitude modulation (QAM) constellation and $x_i$ denote the $i^{th}$ interferer symbol also belonging a unit energy QAM constellation. The received vector is given by
\begin{equation}
\vspace{-.25cm}
\vec{y}=\mathbf{c}x+ \sum_{i=1}^{N_I}\mathbf{c}_ix_i + \mathbf{n},
\end{equation}
where $\mathbf{n}$ is the $N_R \times 1$ additive white complex Gaussian noise vector, with a power of $\sigma^2$ per dimension i.e., $\mathbf{n} \sim \mathcal{CN}(\mathbf{0},\sigma^2 \mathbf{I}_{N_R})$. The interferer channels are modeled as i.i.d. Rician i.e., $\mathbf{c}_i  \sim \mathcal{CN}(\sqrt{a'}\mathbf{m'_i}, b'\mathbf{I}_{N_R})$, where $a'=\frac{\kappa_i}{\kappa_i+1}$, $b'=\frac{1}{\kappa_i+1}$,  $\kappa_i$ is the ratio of the power of the line of sight component to the scattering component of the interferer signals and $\mathbf{m'_i}$ is an $N_R \times 1$ arbitrary vector with elements of unit magnitude. The user channel is also assumed to be i.i.d Rician i.e., $\mathbf{c}  \sim \mathcal{CN}(\sqrt{a}\mathbf{m},b\mathbf{I}_{N_R})$, where $a=\frac{\kappa_s}{\kappa_s+1}$, $b=\frac{1}{\kappa_s+1}$. Note that, the Rician parameter $\kappa_s$ is the ratio of the power of the line of sight component to that of the scattering component and $\mathbf{m}$ is an $N_R \times 1$ mean vector with elements of unit magnitude and uniform phase. Let $E_I''=N_I\times E_I'$ denote the total energy of the interfering signals, where $E_I'$ is the mean energy of each of the interfering signals.
The covariance matrix of the interference term plus the noise term is given by
\begin{equation} \label{R}
 \mathbf{R}=E_I'\mathbf{C'C'}^H + \sigma^2\mathbf{I}=E_I\mathbf{CC}^H + \sigma^2\mathbf{I},
 \end{equation}
 where  $\mathbf{C'}=[\mathbf{ {c}}_1,...,\mathbf{ {c}}_{N_I}]$, $\mathbf{C'} \sim \mathcal{CN}(\sqrt{a'}\mathbf{M'}, b'\mathbf{I}_{N_R} \otimes \mathbf{I}_{N_I})$. $\mathbf{M'}$ is an arbitrary deterministic matrix obtained by stacking $\mathbf{m'_i}$s, such that, $\mathbf{M'}=[\mathbf{m'_1}, \mathbf{m'_2}, \mathbf{m'_3},...., \mathbf{m'_{N_I}}]$ and $\text{tr}(\mathbf{M'M'}^H)=N_RN_I$. Here, $\mathbf{C} \sim \mathcal{CN}(\mathbf{M}, \mathbf{I}_{N_R} \otimes \mathbf{I}_{N_I})$, $E_I=E_I'\times b'$ and $\mathbf{M}=\sqrt{a'}/\sqrt{b'}\mathbf{M'}=\sqrt{\kappa_i}\mathbf{M'}$.
 The received SINR for the OC is given by \cite{chi1}
 \begin{equation} \label{eta}
\eta=E_D\mathbf{ {c}}^H\mathbf{R}^{-1}\mathbf{ {c}},
 \end{equation}
where $E_D$ is the mean energy of the user signal. In the next two sections, we will detail the procedure to obtain the Laplace transform expressions.
\section{Laplace transform for equal power uncorrelated interferers}
A general expression for the Laplace transform $M_{\eta}(s)$ of SINR $\eta$ can now be obtained from Theorem 1 of \cite{chiric}. We further simplify this expression for the specific case of Rician distribution. Let $n_2=\text{max}(N_R,  N_I)$ and $n_1=\text{min}(N_R,  N_I)$\footnote{According to definition, moment generating function (mgf) should ideally exist in an interval around $0$. But in all works including \cite{chiric}, $M_{\eta}(s)$ exists only for $s <0$. So we believe that calling $M_{\eta}(s)$ Laplace transform is a more appropriate, as Laplace transform can be one-sided unlike mgf.}.
\ifCLASSOPTIONtwocolumn
\begin{align}\label{Mgfini}
\nonumber
M_{\eta}(s)&=(-1)^{N_R}(\sigma^2/E_I)^{(N_R-n_1)}\\
& \mathbb{E}_{\boldsymbol{\Lambda}_R} \Bigg[\Bigg(\prod_{i=1}^{n_1}\frac{\sigma^2/E_I+\lambda_i}{\lambda_i^{(N_R-n_1)}}\Bigg)
 \frac{|\mathbf{J}|}{V_{n_1}(\boldsymbol{\Lambda}_R)}\Bigg],
\end{align}
\else
\begin{equation} \label{Mgfini}
 M_{\eta}(s)=(-1)^{N_R}(\sigma^2/E_I)^{(N_R-n_1)} \mathbb{E}_{\boldsymbol{\Lambda}_R} \Bigg[\Bigg(\prod_{i=1}^{n_1}\frac{\sigma^2/E_I+\lambda_i}{\lambda_i^{(N_R-n_1)}}\Bigg)
 \frac{\text{det}(\mathbf{J})}{V_{n_1}(\boldsymbol{\Lambda}_R)}\Bigg],
\end{equation}
\fi
where $V_{n_1}(\boldsymbol{\Lambda}_R)$ is the determinant of the Vandermonde matrix formed by eigenvalues of non-central Wishart matrix $\mathbf{CC}^H$. $\mathbf{J}$ is an $n_1 \times n_1$ matrix with elements,
\begin{equation}\label{J}
\mathbf{J}_{i,j}=\begin{cases}
                  h_1(s,\lambda_i)-\sum_{t=1}^{N_R-n_1}h_t(s,0)\lambda_i^{t-1} , \:  j=1,\\
                  \lambda_i^{N_R-j} , \: j=2,...,n_1,
                 \end{cases}
\end{equation}
and
\begin{equation} \label{ht}
 h_t(s,x)=\frac{_1F_1(t;N_R;\frac{aN_Rs}{xE_I/E_D+\sigma^2/E_D-bs})}{(bsE_D/E_I-\sigma^2/E_I-x)^t},
\end{equation}
 with the series expansion of $_1F_1(.)$ given by $_1F_1(a;b;z)=\sum_{z=0}^{\infty}\frac{(a)_kz^k}{(b)_kk!}$. Laplace transform is derived in \cite{chiric} for two cases: a) Rician signal with Rayleigh interferers b) Rayleigh signal with Rician interferers. In the former case, (\ref{Mgfini}) is used along with the eigenvalue distribution of central Wishart matrix to arrive at a closed form expression for the Laplace transform. In the latter case, the fact that user signal $\mathbf{c}$ exhibits Rayleigh fading and hence invariant under unitary transformation is exploited to derive a closed form expression for the Laplace transform. To the best of our knowledge, there is no open literature that proves that Rician distribution is invariant under unitary transformation. Therefore, for the case of Rician signals with Rician interferers, we propose to evaluate the expectation in (\ref{Mgfini}), by using the eigenvalue distribution of non-central Wishart matrix and subsequently simplify it by using properties of hypergeometric functions. The joint probability density function (pdf) of ordered eigenvalues $( \lambda_1 \; > \; \lambda_2 \; > \, ... \, > \lambda_{n_1})$ of non-central Wishart matrix is given by \cite{shi},
\begin{equation} \label{pdf}
 f(\lambda_1,...,\lambda_s)=c_1|\boldsymbol{\Upsilon}|\prod_{i<j}^{n_1}(\lambda_i-\lambda_j)\prod_{k=1}^{n_1}\lambda_k^{n_2-n_1}e^{-\lambda_k},
 \end{equation}
where $\boldsymbol{\Upsilon}$ is a $n_1 \times n_1$ matrix whose $(i,j)^{th}$ entry $\forall i=1,...,n_1$ is given by,
\begin{equation*}
 \boldsymbol{\Upsilon}_{i,j}=\begin{cases}
                              _0F_1(n_2-n_1+1;w_j\lambda_i),& j=1,...,L,\\
                              \lambda_i^{n_1-j}\frac{(n_2-n_1)!}{(n_2-j)!},& j=L+1,...,n_1,
                             \end{cases}
\end{equation*}
and
\begin{equation*}
 c_1=\frac{e^{-\text{tr}(\boldsymbol{\Omega})}((n_2-n_1)!)^{-n_1}}{\prod_{i=1}^{n_1-L}(n_1-L-i)!\prod_{i=1}^{L}w_i^{n_1-L}\prod_{i<j}^{L}(w_i-w_j)}.
\end{equation*}
Note that $w_i$s are the ordered $L$ non-zero eigenvalues of the non-centrality matrix, $\boldsymbol{\Omega}=\mathbf{M}^H\mathbf{M}$ and the series expansion of the hypergeometric function $_0F_1(.)$ is given by $_0F_1(b;z)= \sum_{k=0}^{\infty}\frac{z^k}{(b)_kk!}$.
Substituting (\ref{pdf}) in (\ref{Mgfini}),
\ifCLASSOPTIONtwocolumn
\begin{align} \label{MGFinter}
  \nonumber
  M_{\eta}(s)&=c_1(-1)^{N_R}\left(\frac{\sigma^2}{E_I}\right)^{(N_R-n_1)}\\
 & \quad \times  \int_{0}^{\infty}\Bigg(\prod_{i=1}^{n_1}(\sigma^2/E_I+\lambda_i) e^{-\lambda_i}\Bigg)|\mathbf{J}||\boldsymbol{\Upsilon}|d\lambda_1...d\lambda_{n_1}.
\end{align}
\else
\begin{align} \label{MGFinter}
\nonumber
 M_{\eta}(s)&=c_1(-1)^{N_R}(\sigma^2/E_I)^{(N_R-n_1)}
 \int_{0}^{\infty}\Bigg[\Bigg(\prod_{i=1}^{n_1}\frac{\sigma^2/E_I+\lambda_i}{\lambda_i^{(N_R-n_1)}}\Bigg)\frac{\text{det}(\mathbf{J})}{V_{n_1}(\boldsymbol{\Lambda}_R)}\Bigg]
 |\boldsymbol{\Upsilon}|\\
 \nonumber
  & \qquad \qquad \times\prod_{i<j}^{n_1}(\lambda_i-\lambda_j)\prod_{k=1}^{n_1}\lambda_k^{n_2-n_1}e^{-\lambda_k}d\lambda_1...d\lambda_{n_1}\\
 &=(-1)^{N_R}(\sigma^2/E_I)^{(N_R-n_1)} \int_{0}^{\infty}c_1\Bigg(\prod_{i=1}^{n_1}(\sigma^2/E_I+\lambda_i) \times e^{-\lambda_i}\Bigg)|\mathbf{J}||\boldsymbol{\Upsilon}|d\lambda_1...d\lambda_{n_1}.
\end{align}
\fi
Using Theorem 2 in Appendix  of \cite{chimim}, 
we can simplify Laplace transform in (\ref{MGFinter}) to obtain,
\begin{equation} \label{MGFfinal}
 M_{\eta}(s)=c|\mathbf{N}|,
\end{equation}
where
\begin{equation*}
 c=\frac{e^{-\text{tr}(\boldsymbol{\Omega})}((n_2-n_1)!)^{-n_1} (-1)^{N_R}(\sigma^2/E_I)^{(N_R-n_1)}}{\prod_{i=1}^{n_1-L}(n_1-L-i)!\prod_{i=1}^{L}w_i^{n_1-L}\prod_{i<j}^{L}(w_i-w_j)},
\end{equation*}
\ifCLASSOPTIONtwocolumn
\begin{align*}
\mathbf N_{i,j}=\begin{cases}
	    \int_{0}^{\infty}(\frac{\sigma^2}{E_I}+x)e^{-x}x^{n_2-N_R}\,_0F_1(n_2-n_1+1;w_ix)\\
	     \quad  [h_1(s,x)-\sum_{t=1}^{N_R-n_1}h_t(s,0)x^{t-1}]dx,\\ 
	    \qquad \qquad \qquad  \qquad \quad \; j=1,\, i=1,..., L,\\
	    \int_{0}^{\infty}(\frac{\sigma^2}{E_I}+x)e^{-x}x^{n_2-N_R}x^{n_1-i}\frac{(n_2-n_1)!}{(n_2-i)!}\\
	     \quad [h_1(s,x)-\sum_{t=1}^{N_R-n_1}h_t(s,0)x^{t-1}]dx,\\
	      \qquad \qquad \qquad \qquad \quad \; j=1,\, i=L+1,..., n_1,\\
	    \int_{0}^{\infty}(\frac{\sigma^2}{E_I}+x)e^{-x}x^{n_2-N_R}\\
        \quad \,_0F_1(n_2-n_1+1;w_ix)x^{N_R-j}dx,\\
        \qquad \qquad \qquad \qquad \quad j=2,...,n_1 \, i=1,..., L,\\
 	    \int_{0}^{\infty}(\frac{\sigma^2}{E_I}+x)e^{-x}x^{n_2-N_R}x^{n_1-i}\frac{(n_2-n_1)!}{(n_2-i)!}x^{N_R-j}dx,\\
        \qquad \quad   j=2,...,n_1 \, i=L+1,..., n_1.\\
         \end{cases}
\end{align*}
\else
\begin{equation*}
 \mathbf N_{i,j}=\begin{cases}
	    \int_{0}^{\infty}(\frac{\sigma^2}{E_I}+x)e^{-x}x^{n_2-N_R}\,_0F_1(n_2-n_1+1;w_ix)
	     [h_1(s,x)-\sum_{t=1}^{N_R-n_1}h_t(s,0)x^{t-1}]dx,\\ \qquad \qquad \qquad  \qquad \quad \; j=1,\, i=1,..., L,\\
	    \int_{0}^{\infty}(\frac{\sigma^2}{E_I}+x)e^{-x}x^{n_2-N_R}x^{n_1-i}\frac{(n_2-n_1)!}{(n_2-i)!}
	    [h_1(s,x)-\sum_{t=1}^{N_R-n_1}h_t(s,0)x^{t-1}]dx,\\  \qquad \qquad \qquad \qquad \quad \; j=1,\, i=L+1,..., n_1,\\
	    \int_{0}^{\infty}(\frac{\sigma^2}{E_I}+x)e^{-x}x^{n_2-N_R}
 \,_0F_1(n_2-n_1+1;w_ix)x^{N_R-j}dx, \: j=2,...,n_1 \, i=1,..., L,\\
 	    \int_{0}^{\infty}(\frac{\sigma^2}{E_I}+x)e^{-x}x^{n_2-N_R}
 x^{n_1-i}\frac{(n_2-n_1)!}{(n_2-i)!}x^{N_R-j}dx, \quad    j=2,...,n_1 \, i=L+1,..., n_1.\\
         \end{cases}
\end{equation*}
\fi
Further simplification of $\vec N_{i,j}$ is given in Appendix \ref{Nsimp}. The final expression for the entries of $\vec N_{i,j}$ is given in  (\ref{Nfin}) and this can be substituted in  (\ref{MGFfinal}) to obtain the final Laplace transform of SINR. In  (\ref{Nfin}), $p=\sigma^2/E_I$, $q=n_2-n_1+1$, $u=aN_RsE_D/E_I$, $v=bsE_D/E_I-\sigma^2/E_I$. Also, note that the expression for $N_{i,j}$, for  $L=n_1$ and $L=0$, which correspond to all Rician faded interferers and all Rayleigh faded interferers respectively, get significantly simplified. When we do a Laplace expansion of the determinant $|\vec N_{L=0}|$ along the first column and substitute $\zeta_t(k)=\frac{\sigma^2}{E_I}\Gamma(n_1+n_2-N_R+t-k+1)+\Gamma(n_1+n_2-N_R+t-k+2)$, we observe that the expression for $M_{\eta}(s)$ for the Rayleigh interferers case is the same as the one obtained in \cite[Eq. 13]{chiric}.
\ifCLASSOPTIONtwocolumn
\begin{figure*}
\begin{equation}\label{Nfin}
\mathbf N_{i,j}=\begin{cases}
	    \sum_{k=0}^{T_2}\frac{w_i^k}{q_k k!}\Bigg[\sum_{l=0}^{T_1}\frac{u^l}{(N_R)_l}
\bigg[-\Gamma (k+n_2-N_R+2) U(l+1,l-k-n_2+N_R,-v) -p  \Gamma (k+n_2-N_R+1)\\
\quad U(l+1,-k-n_2+N_R+l+1, -v)\bigg]\Bigg]
	    -   \sum_{t=1}^{N_R-n_1}\frac{_1F_1(t;N_R;u/v)}{v^t} \bigg[p\Gamma(t+n_2-N_R) \,_1F_1(t+n_2-N_R;q;w_i) \, \\
\quad  +  \Gamma(t+n_2-N_R+1)
_1F_1(t+n_2-N_R+1;q;w_i)\bigg],   \qquad \qquad \qquad \qquad \qquad \qquad \quad  j=1, \, i=1,..., L,\\
\frac{(n_2-n_1)!}{(n_2-i)!}\Bigg[\sum_{l=0}^{T_1}\frac{u^l}{(N_R)_l}
\bigg[-\Gamma(n_2-N_R+n_1-i+2) U(l+1,l-n_2+N_R-n_1+i,-v)\\ 
\quad -p\Gamma(n_2-N_R+n_1-i+1)  U(l+1,-n_2+N_R-n_1+i+l+1, -v)\bigg] - \;  \sum_{t=1}^{N_R-n_1}\frac{_1F_1(t;N_R;u/v)}{v^t} \\
	    \, \bigg[p\Gamma(t+n_2+n_1-N_R-i)  +  \Gamma(t+n_2+n_1-N_R-i+1)\bigg]\Bigg], \qquad  j=1,\, i=L+1,..., n_1,\\
	   p\,_1F_1(n_2-j+1;q;w_i) \Gamma(n_2-j+1) + \Gamma(n_2-j+2)
_1F_1(n_2-j+2;q;w_i) ,  \quad j=2,...,n_1 \, i=1,..., L.\\
 \frac{(n_2-n_1)!}{(n_2-i)!}[p \Gamma(n_2+n_1-i-j+1)
+ \Gamma(n_2+n_1-i-j+2)], \qquad \qquad \qquad   j=2,...,n_1 \, i=L+1,..., n_1.
         \end{cases}
\end{equation}
\end{figure*}
\else
\begin{figure*}
\begin{equation}\label{Nfin}
\mathbf N_{i,j}=\begin{cases}
	    \sum_{k=0}^{T_2}\frac{w_i^k}{q_k k!}\Bigg[\sum_{l=0}^{T_1}\frac{u^l}{(N_R)_l}
\bigg[-\Gamma (k+n_2-N_R+2) U(l+1,l-k-n_2+N_R,-v) \\-p  \Gamma (k+n_2-N_R+1)U(l+1,-k-n_2+N_R+l+1, -v)\bigg]\Bigg]\\
	    - \;  \sum_{t=1}^{N_R-n_1}\frac{_1F_1(t;N_R;u/v)}{v^t}
\; \times \bigg[p\Gamma(t+n_2-N_R)
 \,_1F_1(t+n_2-N_R;q;w_i) \, \\
\quad \; + \, \Gamma(t+n_2-N_R+1)
_1F_1(t+n_2-N_R+1;q;w_i)\bigg],\\  \qquad \qquad \qquad \qquad \qquad \qquad \qquad \qquad \qquad \quad j=1, \, i=1,..., L,\\
\frac{(n_2-n_1)!}{(n_2-i)!}\Bigg[\sum_{l=0}^{T_1}\frac{u^l}{(N_R)_l}
\bigg[-\Gamma(n_2-N_R+n_1-i+2)\\ U(l+1,l-n_2+N_R-n_1+i,-v)\\ -p\Gamma(n_2-N_R+n_1-i+1)  U(l+1,-n_2+N_R-n_1+i+l+1, -v)\bigg]\\
	    - \;  \sum_{t=1}^{N_R-n_1}\frac{_1F_1(t;N_R;u/v)}{v^t}\bigg[p\Gamma(t+n_2+n_1-N_R-i) \, \\
\quad \; + \, \Gamma(t+n_2+n_1-N_R-i+1)\bigg]\Bigg],\\     \qquad \qquad \qquad \qquad \qquad \qquad \qquad \qquad \qquad \quad \; j=1,\, i=L+1,..., n_1,\\
	   p\,_1F_1(n_2-j+1;q;w_i) \Gamma(n_2-j+1) + \Gamma(n_2-j+2)
_1F_1(n_2-j+2;q;w_i) , \\  \qquad \qquad \qquad \qquad \qquad \qquad \qquad \qquad \qquad \quad j=2,...,n_1 \, i=1,..., L.\\
 \frac{(n_2-n_1)!}{(n_2-i)!}[p \Gamma(n_2+n_1-i-j+1)
+ \Gamma(n_2+n_1-i-j+2)], \\ \qquad \qquad \qquad \qquad \qquad \qquad \qquad \qquad \qquad \quad  j=2,...,n_1 \, i=L+1,..., n_1.
         \end{cases}
\end{equation}
\end{figure*}
\fi
\subsection*{Interference-limited scenario}
Recently, there has been a lot of interest in characterizing the performance of cellular networks/wireless system in an interference-limited scenario. Throughput and rate have been studied in \cite{paris, coverage_kmu,outage_kmu, zhou1, chayawan} and references therein, assuming the noise can be neglected (i.e., $\sigma^2=0$), in an interference-limited scenario. Motivated by these works, we now show that the Laplace expansion can be substantially simplified in the case of Rayleigh interferers. Note that an interference-limited scenario is possible only for $N_I > N_R$. For $N_R > N_I$ and $\sigma^2=0$, the receive antennas can cancel every interfering signal. If the number of non-zero eigenvalues $L=0$ as is the case for Rayleigh fading, then
$ M_{\eta}(s)=c|\vec N_{\sigma^2=0, L=0} |$, where $ c=\frac{((n_2-n_1)!)^{-n_1}}{ \prod_{i=1}^{n_1}(n_1-i)!} (-1)^{N_R}\left(\frac{\sigma^2}{E_I}\right)^{(N_R-n_1)}$. 
Note that in $c$, we do not neglect $\sigma^2$.
The expression can be further simplified as shown in Appendix \ref{zero} to obtain,
\ifCLASSOPTIONtwocolumn
\begin{align}\label{laplaceint}
\nonumber
 M_{\eta}(s)&=\frac{  (-1)^{N_R+n_1-1}(\sigma^2/E_I)^{(N_R-n_1)}n_2! } {  (n_2-n_1)! \prod_{i=1}^{n_1}(n_1-i)!}\\
 & \quad
\sum_{i=1}^{n_1}(-1)^{i+1}A(i)\frac{\prod_{j=1}^{n_1}(j-1)!}{(n_1-i)!(i-1)!},
\end{align}
\else
\begin{equation} 
 M_{\eta}(s)=\frac{  (-1)^{N_R+1}(\sigma^2/E_I)^{(N_R-n_1)}n_2! } {  (n_2-n_1)! \prod_{i=1}^{n_1}(n_1-i)!}\sum_{i=1}^{n_1}(-1)^{i+1}A(i)\frac{\prod_{j=1}^{n_1}(j-1)!}{(n_1-i)!(i-1)!},
\end{equation}
\fi
where $A(i)$ is given in Appendix  \ref{zero} below (\ref{Ninitial}).
\section{Laplace Transform for correlated interferers and unequal power interferers}
In the previous section, Laplace transform expressions are derived for the case of equal power uncorrelated interferers. But, in practice, the interferers can have different power and/or can be correlated.  In a practical cellular system, there can be one or more of the following: a) receiver side correlation, b) interferer correlation, c) unequal power interferers.
\par The general non-central Wishart matrix $\vec W$ is written as $\vec W=\vec{C'C'}^H$, where $\vec C' \sim \mathcal{CN}(\vec M, \vec \Sigma \otimes \vec \Psi)$. Here, the $N_R \times N_R$ matrix $\vec \Sigma$ denotes the receive correlation and the $N_I \times N_I$ matrix $\vec \Psi$ denotes the transmit correlation or interferer correlation in our case. Suppose, we consider only receive side correlation and assume that the interferer correlation is not present, i.e., $\vec \Psi$ is an identity matrix. This reduces to the non-central Wishart matrix denoted by $\vec W \sim \mathcal{CW}(N_I, \vec \Sigma , \vec \Sigma^{-1}\vec{MM}^H)$.  This case, where $\vec \Psi$ is assumed to be an identity matrix, is widely discussed in the literature. The eigenvalue distribution of this case, i.e., a non-central Wishart matrix with a covariance matrix $\vec \Sigma$ which is not an identity matrix, is analyzed in \cite{ratnarajah_2005} in terms of zonal polynomials. However, using this eigenvalue distribution to obtain the Laplace transform expression of $\eta$ becomes mathematically intractable. Hence, considering receive correlation is beyond the scope of this work. 
\par On the other hand, the cases of $\vec \Psi$ being a diagonal matrix, i.e., unequal power interferers or $\vec \Psi$ being a full matrix, i.e.,  correlated interferers, have barely received attention in statistic literature. There do not even exist matrix variate and eigenvalue distribution results for this case. However, we do provide results for this case by considering the problem as two sub-problems a) for $N_R \geq N_I$ exact results are provided, b) for $N_R < N_I$ approximate results are provided.  In short, in this section, we derive Laplace transform expressions for $\vec C' \sim \mathcal{CN}( \vec M , \vec I_{N_R} \otimes \vec \Psi)$. Note, $\vec W = \vec{C'C'}^H$ can be decomposed into  $\vec W=\vec{ C \vec \Psi  C}^H$, such that $\vec C \sim \mathcal{CN}( \vec M \vec \Psi ^{-\frac{1}{2} }, \vec I_{N_R} \otimes \vec I_{N_I})$ \cite{mckay_capacity}. Let us first consider the case of correlated interferers. The covariance matrix of the interference term plus the noise term is given by
\begin{equation} \label{R1}
 \mathbf{R}=\mathbf{C \vec \Psi C}^H + \sigma^2\mathbf{I}.
 \end{equation}
  The received SINR for the OC is given by (\ref{eta}). We will consider this problem as two cases: a) $N_R \geq N_I$, b) $N_R < N_I$. 
\subsection{$N_R \geq N_I$}
Similar to (\ref{Mgfini}), the general expression for the Laplace transform $M_{\eta}(s)$ of SINR $\eta$ is given by,
\ifCLASSOPTIONtwocolumn
\begin{align}\label{MGFini1}
\nonumber
M_{\eta}(s)&=(-1)^{N_R}(\sigma^2)^{(N_R-n_1)}\\
& \mathbb{E}_{\boldsymbol{\Lambda}_R} \Bigg[\Bigg(\prod_{i=1}^{n_1}\frac{\sigma^2+\lambda_i}{\lambda_i^{(N_R-n_1)}}\Bigg)
 \frac{|\mathbf{J}|}{V_{n_1}(\boldsymbol{\Lambda}_R)}\Bigg],
\end{align}
\else
\begin{equation}\label{MGFini1}
 M_{\eta}(s)=(-1)^{N_R}(\sigma^2)^{(N_R-n_1)} \mathbb{E}_{\boldsymbol{\Lambda}_R} \Bigg[\Bigg(\prod_{i=1}^{n_1}\frac{\sigma^2+\lambda_i}{\lambda_i^{(N_R-n_1)}}\Bigg)
 \frac{\text{det}(\mathbf{J})}{V_{n_1}(\boldsymbol{\Lambda}_R)}\Bigg],
\end{equation}
\fi
where $V_{n_1}(\boldsymbol{\Lambda}_R)$ is the determinant of the Vandermonde matrix formed by eigenvalues of non-central Wishart matrix $\mathbf{C \Psi C}^H$. $\mathbf{J}$ is an $n_1 \times n_1$ matrix with elements given by (\ref{J}) and
\begin{equation}
 h_t(s,x)=\frac{_1F_1(t;N_R;\frac{aN_Rs}{x/E_D+\sigma^2/E_D-bs})}{(bsE_D-\sigma^2-x)^t}.
\end{equation}
Recall that, in the case of equal power uncorrelated interferers, we simplified the expression in (\ref{Mgfini}) using the eigenvalue distribution of the non-central Wishart matrix. But for the case of correlated interferers, there exists no matrix variate distribution formula in the open literature and deriving one requires integration over the Stiefel manifold \cite{ratnarajah_2005}. Also, there exists no eigenvalue distribution for this case. Hence, initially, we consider the case of Rayleigh-faded interferers, i.e., $\vec M= \vec 0$.
\subsubsection{Rayleigh faded correlated interferers} 
We exploit the property that $\vec W=\vec{ C \vec \Psi  C}^H$ has the same non-zero eigenvalues as that of $\vec \Psi^{\frac{1}{2}}\vec C^H \vec C\vec \Psi^{\frac{1}{2}}$, where $\vec C^H \vec C$ is also a Wishart matrix. The eigenvalue distribution of $\vec \Psi^{\frac{1}{2}}\vec C^H \vec C\vec \Psi^{\frac{1}{2}}$, is given by \cite{random_book},
\begin{equation} \label{pdf1}
 f(\lambda_1,...,\lambda_s)=c_1|\boldsymbol{\Upsilon}|\prod_{i<j}^{N_I}(\lambda_i-\lambda_j)\prod_{k=1}^{N_I}\lambda_k^{N_R-N_I},
 \end{equation}
where $\boldsymbol{\Upsilon}$ is a $N_I \times N_I$ matrix whose $(i,j)^{th}$ entry $\forall i,j=1,...,N_I$ is given by,
$\boldsymbol{\Upsilon}_{i,j}=e^{-\frac{\lambda_i}{r_j}}$ and $
 c_1=(-1)^{\frac{1}{2}N_I(N_I-1)}\frac{|\vec \Psi |^{-N_R}}{\prod_{i<j}^{N_I}(\frac{1}{r_i}-\frac{1}{r_j}) \prod_{k=1}^{N_I}(N_R-k)!}$. Also, note that $r_i$s are the ordered $N_I$ distinct non-zero eigenvalues of $\vec \Psi $. Using (\ref{pdf1}) to simplify (\ref{MGFini1}), we obtain the Laplace transform of $\eta$ as
\begin{align} \label{MGFinter1}
\nonumber
 M_{\eta}(s) &=(-1)^{N_R}(\sigma^2)^{(N_R-N_I)}\\
 & \times \int_{0}^{\infty}c_1\prod_{i=1}^{n_1}(\sigma^2+\lambda_i) |\mathbf{J}||\boldsymbol{\Upsilon}|d\lambda_1...d\lambda_{N_I}.
\end{align}
As in the case of equal power interferers, we use Theorem 2 in the Appendix  of \cite{chimim} and the identity $ \int_{0}^{\infty}e^{-px}x^{s-1}dx =p^{-s}\Gamma(s)$ from \cite{int}. We can solve the integral for $j=1$, by expanding the $\,_1F_1$ hypergeometric series and interchanging the integration and summation. The approach followed in Appendix \ref{Nsimp} can be followed here.
The Laplace transform after simplification becomes, 
\begin{equation} \label{MGFfinal1}
 M_{\eta}(s)=c|\mathbf{N}|,
\end{equation}
where $c= \frac{ (-1)^{N_R}(\sigma^2)^{(N_R-N_I)}(-1)^{\frac{1}{2}N_I(N_I-1)} |\vec \Psi |^{-N_R}}{\prod_{i<j}^{N_I}(\frac{1}{r_i}-\frac{1}{r_j}) \prod_{k=1}^{N_I}(N_R-k)!}$
and $N$ is given by
\ifCLASSOPTIONtwocolumn
\begin{equation}\label{Nunequal}
 \mathbf N_{i,j}=\begin{cases}
	    \sum_{l=0}^{\infty}\frac{(aN_RsE_D)^l}{(N_R)_l}(\sigma^2|bsE_D-\sigma^2|^{-l}\\
	    \: U(1, 1-l, r_i|bsE_D-\sigma^2|) + |bsE_D-\sigma^2|^{-l+1}\\
	    \quad \quad U(2, 2-l, r_i|bsE_D-\sigma^2|))\\
	    \quad -\sum_{t=1}^{N_R-N_I}\frac{_1F_1(t;N_R;\frac{aN_Rs}{\sigma^2/E_D-bs})}{(bsE_D-\sigma^2)^t} (\sigma^2 r_i^t \Gamma(t)\\
	     \quad \quad + r_i^{t+1}\Gamma(t+1)], \quad  j=1,\, i=1,..., N_I,\\
 	   \sigma^2 r_i^{N_R-j+1}\Gamma(N_R-j+1)+ r_i^{N_R-j+2} \\
 	    \quad \Gamma(N_R-j+2),  j=2,...,N_I \, i=1,..., N_I,
         \end{cases}
\end{equation} 
\else
\begin{equation}\label{Nunequal}
 \mathbf N_{i,j}=\begin{cases}
	    \sum_{l=0}^{\infty}\frac{(aN_RsE_D)^l}{(N_R)_l}(\sigma^2|bsE_D-\sigma^2|^{-l}U(1, 1-l, r_i|bsE_D-\sigma^2|)\\
	    \quad \quad + |bsE_D-\sigma^2|^{-l+1}U(2, 2-l, r_i|bsE_D-\sigma^2|))\\
	    \quad \quad -\sum_{t=1}^{N_R-N_I}\frac{_1F_1(t;N_R;\frac{aN_Rs}{\sigma^2/E_D-bs})}{(bsE_D-\sigma^2)^t} (\sigma^2 r_i^t \Gamma(t)+ r_i^{t+1}\Gamma(t+1)], \quad  j=1,\, i=1,..., N_I,\\
 	   \sigma^2 r_i^{N_R-j+1}\Gamma(N_R-j+1)+ r_i^{N_R-j+2}\Gamma(N_R-j+2), \quad   j=2,...,N_I \, i=1,..., N_I,\\
         \end{cases}
\end{equation} 
\fi
where $r_i$, $1 \leq i \leq N_I$ are the eigenvalues of $\vec \Psi$. We can truncate the converging infinite series for $j=1$ at a finite value, with an arbitrarily small truncation error. The convergence proof is similar to the one given in Appendix \ref{Nsimp}. Note that this is an exact Laplace transform expression and is novel for the case of Rayleigh faded correlated interferers with Rician faded users. Earlier works like \cite{chiric}, considers only equal power uncorrelated interferers, while recent works like \cite{sri} consider only Rayleigh faded user. An approximation which works for $\sigma^2 \approx 0$ is also derived in Appendix \ref{unequal}. The expression is as follows:
\begin{equation} \label{mgfunequal}
 M_{\eta}(s) \approx c \sum_{i=1}^{N_I}(-1)^{i+1}B(i)r_i^{-N_R+N_I-2}|V^{i}(\vec r)|,
\end{equation}
where $ c=\frac{(-1)^{N_R}(\sigma^2)^{(N_R-N_I)}(-1)^{\frac{1}{2}N_I(N_I-1)}|\vec \Psi |^{-N_R}}{\prod_{i<j}^{N_I}(\frac{1}{r_i}-\frac{1}{r_j}) \prod_{k=1}^{N_I}(N_R-k)!}$, $A(i)$ and $|V^{i}(\vec r)|$ are given in Appendix \ref{unequal}.
\subsubsection{Rician faded correlated interferers} For the case of Rician faded interferers, the above approach is not possible even for $N_R \geq N_I$. This is because we have to use the zonal-polynomial based eigenvalue distribution from \cite{ratnarajah_2005} to simplify the Laplace transform, which is not mathematically tractable.
Nevertheless, we arrive at a mathematically tractable solution, wherein we propose to approximate the non-central Wishart matrix by a central Wishart matrix through moment matching and then use the derived expression given in (\ref{MGFfinal1}).
 \par The expected value of any matrix of the form $\vec A^H \vec A $, if $\vec A \sim \mathcal{CN}(\vec M, \vec I_{N_I} \otimes \vec I_{N_R})$, with $N_R$ degrees of freedom, is given by $E[\vec A^H \vec A]=N_R+\vec M^H\vec M$. This implies that for $\vec W= \vec \Psi^{\frac{1}{2}}\vec C^H \vec C \vec \Psi^{\frac{1}{2}}$, the first moment is given by, $E[\vec W]=N_R\vec \Psi+\vec  \Psi^{1/2} \vec M^H \vec M \vec \Psi^{1/2}$. The first moment of any central Wishart matrix $\vec W_1 \sim \mathcal{CW}(N_R, \vec \Phi)$ with the same degree of freedom $N_R$, is given by $E[\vec W_1]=N_R \vec \Phi$. When the first moments of the $\vec W $ and $\vec W_1$ are equated, we obtain
$\vec \Phi = \vec \Psi+\frac{1}{N_R}\vec  \Psi^{1/2} \vec M^H \vec M \vec \Psi^{1/2}$.
We have now obtained a central Wishart approximation of the non-central Wishart matrix. 
Hence, $\vec W$ can be approximated as a central Wishart $\mathcal{CW}(N_R, \vec \Psi  + \frac{1}{N_R}\vec \Psi ^{1/2}\vec M^H\vec M\vec \Psi ^{1/2})$. A similar approximation is performed in \cite{mimo_rician}. Now that we have a central Wishart matrix, the expressions derived for the case of Rayleigh faded interferers hold, but with the matrix $\vec \Psi $ in (\ref{MGFfinal1}) replaced by $\vec \Psi  + \frac{1}{N_R}\vec \Psi ^{1/2}\vec M^H\vec M\vec \Psi ^{1/2}$. 
\subsubsection{Unequal power interferers}
\par All the above analysis holds for a general $\vec \Psi$. For the case of unequal power interferers, $\vec \Psi$ is just a diagonal matrix, with the interferer powers occupying the diagonal.
Hence, the Laplace transform expressions (\ref{MGFfinal1}) can be used for unequal power Rayleigh-faded and Rician-faded interferers respectively.
 \subsection{ Rayleigh faded correlated interferers for $N_I > N_R$}
 For the case of Rayleigh-faded correlated interferers, for $N_I > N_R$, the covariance matrix of the interference term plus the noise term is given by
 $\mathbf{R}=\mathbf{C \vec \Psi C}^H + \sigma^2\mathbf{I}$.
 Here, $\mathbf{C} \sim \mathcal{CN}(\mathbf{0}, \mathbf{I}_{N_R} \otimes \mathbf{I}_{N_I})$. From \cite{ohlson}, the distribution of $\vec W= \mathbf{C \vec \Psi C}^H$ is the same as that of $\sum_{i=1}^{N_I} \lambda_i \vec W_i$, where $\lambda_i$ are the eigenvalues of $\vec \Psi$ and $\vec W_i \sim \mathcal{CW}(1, \vec I_{N_R})$. Though this method works for $N_R \geq N_I$, we can use the analysis given in the previous subsection for determining the Laplace expansion.
\par From \cite{kumar_exact}, the sum of central Wishart matrices can be approximated by another central Wishart matrix. In our case, from \cite{kumar_exact}, $\vec W \approx \vec S \frac{\sum_{i=1}^{N_I}\lambda_i}{p_s}$, where $\vec S \sim \mathcal{CW}(p_s, \vec I_{N_R})$ and $p_s= \Bigg[ \frac{(\sum_{i=1}^{N_I}\lambda_i)^2}{\sum_{i=1}^{N_I}\lambda_i^2} \Bigg]$ rounded to the nearest integer. Note that, this has reduced to a case of a Wishart matrix with an identity covariance matrix. Hence, the Laplace transform expression derived for the case of equal power Rayleigh interferers, i.e., expressions corresponding to $L=0$ given in Section III, can now be used. Also, the determinant simplification that has been derived in the case of equal power Rayleigh faded interferers holds for this case. 
 \section{SER Expressions}
Using the standard assumption that the contribution of the interference and the noise at the output of optimum combiner, for a fixed $\eta$, can be well-approximated to be Gaussian as in \cite{poor} and \cite{chifinter} and references therein, the probability of symbol error for an M-ary square QAM constellation is given by \cite{proakis},
\begin{equation} \label{peini}
\mathcal{P}_e \approx k_1Q(\sqrt{k_2\eta})-k_3Q(\sqrt{k_2\eta})^2,
\end{equation}
 where $k_1=4\big(1-\frac{1}{\sqrt M}\big)$, $k_2=\frac{3}{M-1}$, $k_3=\frac{k_1^2}{4}$ and the Q-function is given by $Q(x)=\frac{1}{2\pi}\int_{x}^{\infty}e^{-u^2/2}du$.
 The assumption that the contribution of the interference and the noise at the output of OC, for a fixed $\eta$, is Gaussian, is valid even when the number of interferers $N_I$ is small \cite{chifinter} and such a system model assumption is made in a number of papers \cite{chi1}, \cite{chiinter2}, \cite{chiinter3} to derive the SER expression.
Using the approximation $\textstyle Q(x) \approx  \textstyle \frac{1}{12} e^{-\frac{1}{2}x^2} +\frac{1}{4} e^{-\frac{2}{3}x^2}$, one can write $\mathcal{P}_e$ as \cite{bounds}, $\mathcal{P}_e \approx \sum_{l=1}^{5} a_l e^{-b_l\eta}$,
where $a_1$=$\frac{k_1}{12}$, $a_2$=$\frac{k_1}{4}$, $a_3$=$\frac{-k_3}{144}$, $a_4$=$\frac{-k_3}{16}$, $a_5$=$\frac{-k_3}{24}$, $b_1$=$\frac{k_2}{2}$, $b_2$=$\frac{2k_2}{3}$, $b_3$=${k_2}$, $b_4$=$\frac{4k_2}{3}$ and $b_5$=$\frac{7k_2}{6}$.
The average SER obtained by averaging $\mathcal{P}_e$ over all channel realizations is,
\begin{align} \label{pe}
SER \approx \mathbb{E}_{\eta}[\mathcal{P}_e]
=\mathbb{E}_{\eta}[\sum_{l=1}^{5} a_l e^{-b_l\eta}]
=\sum_{l=1}^{5} a_l M_{\eta}(s)|_{s=-b_l},
\end{align}
where $M_{\eta}(s)$ is the Laplace transform of $\eta$ and $\mathbb{E}_{\mathbf{\eta}}$ denotes expectation over SINR $\eta$. Now, the SER approximations can be directly obtained by substituting the Laplace transform expressions in the above equation. Wherever the Laplace transform is simplified to circumvent the determinant expansion, we get simplified SER expressions. For example, for an interference-limited scenario ($N_I > N_R$ and $\sigma^2=0$), by substituting $n_1=N_R$ and $n_2=N_I$,
the SER becomes,
\ifCLASSOPTIONtwocolumn
\begin{align}\label{highSNR1}
\nonumber
SER &\approx \sum_{l=1}^{5} a_l \frac{  (-1)^{N_R+1}N_I! } {  (N_I-N_R)! \prod_{i=1}^{N_R}(N_R-i)!} \\
& \times \sum_{i=1}^{N_R}(-1)^{i+1}A(i)\frac{\prod_{j=1}^{N_R}(j-1)!}{(N_R-i)!(i-1)!}|_{s=-b_l}.
\end{align}
\else
\begin{equation}\label{highSNR1}
SER \approx \sum_{l=1}^{5} a_l \frac{  (-1)^{N_R+1}N_I! } {  (N_I-N_R)! \prod_{i=1}^{N_R}(N_R-i)!}\sum_{i=1}^{N_R}(-1)^{i+1}A(i)\frac{\prod_{j=1}^{N_R}(j-1)!}{(N_R-i)!(i-1)!}|_{s=-b_l}.
\end{equation}
\fi
Ours is the first work to obtain SER expression in an interference-limited scenario, for Rayleigh faded interferers in a closed form. All existing works, so far, require an explicit evaluation of the determinant. Further, the expression derived also gives an approximation of SER, for $N_I > N_R$ for very low noise values $\sigma^2 \approx 0$. If we substitute $n_1= N_I$ and $n_2=N_R$ in (\ref{MGFfinal}) and ignore the $\sigma^2$ term inside the determinant, we also obtain an approximation for the SER as,
\ifCLASSOPTIONtwocolumn
\begin{align}\label{highSNR2}
\nonumber
SER &\approx \sum_{l=1}^{5} a_l \frac{  (-1)^{N_R+1}(\sigma^2/E_I)^{(N_R-N_I)}N_R! } {  (N_R-N_I)! \prod_{i=1}^{N_I}(N_I-i)!}\\
&\times \sum_{i=1}^{N_I}(-1)^{i+1}A(i)\frac{\prod_{j=1}^{N_I}(j-1)!}{(N_I-i)!(i-1)!}|_{s=-b_l}.
\end{align}
\else
\begin{equation}\label{highSNR2}
SER \approx \sum_{l=1}^{5} a_l \frac{  (-1)^{N_R+1}(\sigma^2/E_I)^{(N_R-N_I)}N_R! } {  (N_R-N_I)! \prod_{i=1}^{N_I}(N_I-i)!}\sum_{i=1}^{N_I}(-1)^{i+1}A(i)\frac{\prod_{j=1}^{N_I}(j-1)!}{(N_I-i)!(i-1)!}|_{s=-b_l}.
\end{equation}
\fi
Note that the dependence of $c$ term on $\sigma^2$ is not present for $N_I > N_R$. On the other hand, the $\sigma^2$ term exists in $c$ term for $N_R > N_I$.
Note that SER expressions obtained for Rayleigh interferers in \cite{chiric} involve not only an explicit evaluation of determinants but also numerical integration, while results here require neither.
For the case of Rayleigh/Rician faded correlated interferers and $N_R > N_I$, we can substitute (\ref{MGFfinal1}) in (\ref{pe}) to get the SER approximations. For this case also, we have a Laplace expansion in (\ref{mgfunequal}) that doesn't involve determinant evaluation. This approximation works very well when $\sigma^2$ is actually small or when interferer powers are large compared to $\sigma^2$. This can be further substituted in (\ref{pe}), to obtain the approximate expression for SER as,
\begin{align}\label{highSNRunequal}
\nonumber
SER &\approx \sum_{l=1}^{5} \frac{ a_l (-1)^{N_R}(\sigma^2)^{(N_R-N_I)}(-1)^{\frac{1}{2}N_I(N_I-1)}|\vec \Psi |^{-N_R}}{\prod_{i<j}^{N_I}(\frac{1}{r_i}-\frac{1}{r_j}) \prod_{k=1}^{N_I}(N_R-k)!}\\
& \times \sum_{i=1}^{N_I}(-1)^{i+1}B(i)_{s=-b_l}r_i^{-N_R+N_I-2}|V^{i}(\vec r)|.
\end{align}
In case of correlated or unequal power Rayleigh faded interferers and $N_I > N_R$, we have discussed an approximate expression for Laplace transform in Section III.B, which can be used to determine the SER expressions. In case, of correlated or unequal power Rician faded interferers for $N_I > N_R$, it is mathematically intractable to give an Laplace transform expression and hence derive an SER expression. Nevertheless, the existing SER expressions derived for the case of equal power uncorrelated interferers can be used as an upper bound. If we consider all the interferers to have the same power as that of the maximum-power interferer, our expression gives an upper bound for the actual SER, i.e., our expressions give the worst case SER.  Similarly, the  expressions for the uncorrelated case gives the worst case SER, i.e., a good upper bound on for the actual SER of correlated interferers. This is because, correlated interferers cause partial interference alignment \cite{sri} and hence the receive antennas can cancel the interferers better, leading to a lower SER when compared to the uncorrelated case.

 \section{Outage and Rate approximations}
Apart from SER, outage probability and rate are the other performance metrics that are useful in characterizing the performance of any wireless system . In the preceding section, we used the exponential approximation to determine the approximate SER at the output of OC. But, no such straightforward method exists for determining expressions for outage probability and rate. This is so because expressions for the pdf of the SINR or signal to interference ratio (SIR) are mathematically intractable to derive. However, we, in this section, detail a moment-matching method to determine approximate expressions for outage probability and rate. To the best of our knowledge, ours is the first work to obtain even approximate expressions for outage probability and rate. For doing this, we will first determine the exact moments of the SINR.
\subsection{Moments of Rician-Rician}
The $l^{th}$ moment of SINR for Rician faded user and Rician faded interferers is given by, 
\begin{align*}
\mu_l^{Ric-Ric}&=\frac{d^l}{ds^l}M_{\eta}(s)|_{s=0}=\alpha_{l}^{Ric}c\sum_{k=1}^{n_1}(-1)^{k}|Y_k|d_l\\ 
\end{align*} 
where $c=\frac{e^{-\text{tr}(\boldsymbol{\Omega})}((n_2-n_1)!)^{-n_1}}{\prod_{i<j}^{n_1}(w_i-w_j)}(-1)^{N_R}(\sigma^2/E_I)^{(N_R-n_1)}$, $d_l$ is given by (\ref{dl}), $Y_k$ is the matrix formed by omitting the $k^{th}$ row and $1^{st}$ column of the matrix $Y_{i,j}=  \frac{\sigma^2}{E_I}\,_1F_1(n_2-j+1;n_2-n_1+1;w_k) \Gamma(n_2-j+1)+  \Gamma(n_2-j+2)
_1F_1(n_2-j+2;n_2-n_1+1;w_i)$ and $\alpha_l^{Ric}= b^l \sum_{k=0}^{l}{ l \choose k} \frac{(aN_R/b)^k}{(N_R)_k}$. The derivation is given in Appendix \ref{moments} and is very similar to the one given in \cite{chiric}. Though we have derived the moments for the uncorrelated case in the presence of noise, similar moment expressions can be obtained for an interference-limited scenario or for Rayleigh faded correlated/unequal power interferers. These derivations are not given here due to space constraints. 
\subsection{Moment matched approximation}
Inspired by the simplicity of the results in \cite{shah}, we now use the idea of moment matching to approximate the SINR/SIR random variables for our case too. If we had Rayleigh faded users and interferers as in the case of \cite{shah}, the SIR would be distributed according to a beta-prime distribution, because the ratio of two gamma random variables follow a beta-prime distribution. We propose to match the first two moments of the SINR/SIR with the moments of the beta-prime distribution. As an example, let us consider the interference-limited scenario, i.e., $N_I > N_R$ and $\sigma^2=0$. The $l^{th}$ moment is given by 
\begin{align}
\nonumber
\mu_l &= \frac{e^{-\text{tr}(\boldsymbol{\Omega})}((N_I-N_R)!)^{-N_R}}{\prod_{i<j}^{N_R}(w_i-w_j)}(-1)^{N_R}\sum_{k=1}^{N_R}(-1)^{k+1}\\
\nonumber
&  \times (-1) b^l l! (\frac{E_D}{E_I})^l \sum_{m=0}^{l}{ l \choose m} \frac{(\frac{aN_R}{b})^m \Gamma(N_I-N_R-l+1)}{(N_R)_m}  \\
& \times \, _1F_1(N_I-N_R-l+1, N_I-N_R+1, w_k) |Y_k|,
\end{align}
where $Y_{i,j}= \Gamma(N_I-j+2)
_1F_1(N_I-j+2;N_I-N_R+1;w_i)$ and $Y_k$ is the matrix $Y$ with $k^{th}$ row and first column removed. 
The first two moments $\mu_1$ and $\mu_2$ of the SIR can be matched with the first two moments $\frac{\alpha}{\beta-1}$ and $\frac{\alpha(\alpha+1)}{(\beta-1)(\beta-2)}$, respectively, of a beta prime distribution with parameters $\alpha, \beta$. 
In other words,
$\frac{\alpha}{\beta-1}= \mu_1$ and
$\frac{\alpha(\alpha+1)}{(\beta-1)(\beta-2)}= \mu_2$. This implies that
\begin{equation}
\beta= \frac{\mu_1^2- \mu_1-2\mu_2}{\mu_1^2-\mu_2}
\end{equation}
and
\begin{equation}
\alpha=\mu_1(\beta-1).
\end{equation}
Since the first and second moments do not involve any infinite summations, the parameters $\alpha$ and $\beta$ are obtained in closed form. These parameters can be substituted in the cumulative distribution function (CDF) of a beta-prime distributed random variable, say $Z$, with parameters $\alpha$ and $\beta$ given by \cite[Eq 2]{cordeiro}
\begin{align}\label{approx}
P_0(Z <z) = \frac{(\beta)_{\alpha} z^{\alpha}  \,_2F_1(\alpha+\beta, \alpha, \alpha+1, -z)}{  \Gamma(\alpha+1)}
\end{align} 
to obtain the closed form outage expression for OC. Here, $\,_2F_1(.)$ denotes the Gauss hypergeometric function \cite[15.1.1]{abr}. Similarly using the pdf of beta-prime distribution given by $f(z)=\frac{\Gamma(\alpha+\beta)}{\Gamma(\alpha)\Gamma(\beta)}z^{\alpha-1}(1+z)^{-\alpha-\beta}$, one can obtain the approximate rate as
\begin{align}\label{rate}
\nonumber
R &\approx \int_{0}^{\infty}\text{log}_2(1+x)\frac{\Gamma(\alpha+\beta)}{\Gamma(\alpha)\Gamma(\beta)}x^{\alpha-1}(\alpha+x)^{-\alpha-\beta}dx\\
&=\frac{\psi(\alpha+\beta)-\psi(\beta)}{ln 2},
\end{align}
where $\psi(.)$ is the di-gamma function \cite[6.3.1]{abr}. 
The second equality is obtained by using the integral identity \cite[4.292.14]{int}.\\
Here, we have considered the case of all interferers to be Rician faded. We can even consider the case of a mix of Rayleigh faded and Rician faded interferers, especially single rank non-centrality matrix $\vec{MM}^H$ or the case of correlated interferers and do a similar moment matching to obtain the approximate outage probability and rate expressions.
\subsection{Analysis using stochastic ordering}
Now, we can study the impact of the Rician parameters on the approximate expressions using stochastic ordering. According to \cite[Theorem 1.A.12]{shaked}, for two random variables $X$ and $Y$ with pdf $f$ and $g$, respectively, if 
$\mathcal{S}^-(g-f)=1$ and the sign sequence is $-,+$, then $X \leq_{st} Y$ or in other words $P( X \leq x) \geq P(Y \leq y)$, i.e., outage probability of $X$ is greater than that of $Y$. Here $\mathcal{S}^-$ denotes the number of sign changes and $ \leq_{st}$ denotes stochastic ordering  \cite[1.A.1]{shaked}. In case we assume $f$ and $g$ to be beta-prime with parameters $(\alpha, \beta+ \delta)$ and $(\alpha, \beta)$, we can easily prove that the sign change from $-$ to $+$ occurs at $z= \Big(\frac{(\beta+\delta)_{\alpha}}{(\beta)_{\alpha}}\Big)^{\frac{1}{\delta}}-1$, where $(x)_y$ denotes the Pochhammer symbol \cite[13.1.2]{abr}. This implies that, $P_0$ given by (\ref{approx}), evaluated using parameter $\beta+\delta$ is greater than $P_0$ evaluated using the parameter $\beta$.
 Similarly, we can also prove $P_0$ evaluated using the parameter $\alpha+\delta$ is lesser than $P_0$ evaluated using the parameter $\alpha$. This method is also adopted in \cite{srinivasan_secrecy}. Since, it is intractable to directly analyze the approximate SER expressions, we'll introduce a Laplace ordering result to connect the variations in outage probability with variations in SER. Let $X$ and $Y$ be two non-negative random variables such that
\begin{equation}
\mathbb{E}[exp(sX)]  \geq \mathbb{E}[exp(sY )], \quad  \forall s <  0.
\end{equation}
Then $X$ is said to be smaller than $Y$ in the Laplace transform order denoted by $X \leq_{Lt} Y$. According to \cite[Theorem 5.A.6]{shaked}, $X \leq_{St} Y$, then $X \leq_{Lt} Y$. In other words, $P( X \leq x) \geq P(Y \leq y)$ implies $E[exp(sX)]  \geq E[exp(sY )]$. Recall that in the probability of error is given by (\ref{peini}), if we neglect the second term, which is usually small in magnitude, SER, by exponential approximation of Q-function, is given by 
\begin{equation}
    SER \approx \mathbb{E}_{\eta}[\sum_{l=1}^{2} a_l e^{-b_l\eta}],
\end{equation}
where $a_1$=$\frac{k_1}{12}$, $a_2$=$\frac{k_1}{4}$, $b_1$=$\frac{k_2}{2}$ and $b_2$=$\frac{2k_2}{3}$.
Now if $P( X \leq x) \geq P(Y \leq y)$, then by the preceding Laplace ordering result, SER for $X$ is greater than that for $Y$. Though, these results actually pertain to the moment-matched $\eta$ and not the actual $\eta$, we can still get some approximate trends in SER based on the variations in outage probability. We also can make the following inferences about variation of outage probability and SER with respect to fading parameters: 
\begin{itemize}
\item I1) We can observe that an increase in $E_I$, decreases $\mu_2$ and $\mu_1$, which in turn increases $\beta$. Similarly, an increase in $E_I$ decreases $\alpha$. According to the previous stochastic ordering result, an increase in $\beta$ and a decrease in $\alpha$ increases the outage probability. By the Laplace ordering result, the SER also increases. 
\item I2) It is also clear that an increase in Rician parameter of the user $\kappa$ decreases $\mu_2$, keeping $\mu_1$ constant. This in turn increases $\alpha$. Also, $\beta$ increases very negligibly. Though, the stochastic ordering result cannot be used directly if both $\alpha$ and $\beta$ increases, the increase in $\beta$ is actually negligible. Hence, the outage probability decreases with increase in $\alpha$ due to an increase in $\kappa$. Also, by the Laplace ordering result, the SER decreases. 
\item  I3) Also, an increase in $N_R$ or a decrease in $N_I$ decreases $\beta$ and increases $\alpha$. This in turn decreases the outage probability.
\end{itemize}
\section{ A short extension to $\kappa-\mu$ faded user and Rayleigh faded interferers}
Recently there has been a lot of interest in studying the effect of a general fading model such as the $\kappa-\mu$ fading model for a LoS scenario. Moreover, $\kappa-\mu$ fading also includes Rician, Nakagami-m, Rayleigh and one-sided Gaussian as its special cases \cite{yacoub}. Given, the fairly complicated nature of these general fading distributions and due to unavailability of a matrix model for these distributions, it is mathematically intractable to determine any expressions considering $\kappa-\mu$ fading in the interferers. However, it is possible to consider $\kappa-\mu$ fading in the user and Rayleigh fading in the interferers and determine exact OP and approximate SER expressions. 
Recall that the SINR for OC is given by \cite{chiric}, 
$\eta=\frac{E_D}{E_I}\vec c^H \vec R^{-1} \vec c$,
where $\vec R = \sum_{i=1}^{N_I}\vec c_i \vec c_i^H$, 
where the user $\vec c$ is $\kappa-\mu$ faded and the interferers $\vec c_i$ are Rayleigh faded. We consider only the case of $N_I > N_R$ and neglect noise in the analysis.  Note that, if noise is neglected, $\vec R$ is a complex Wishart matrix, i.e., $\vec R \sim \mathcal{CW}_{N_R}(N_I, \vec I_{N_R})$. From \cite{nagarexp}, it is known that for any $\vec R \sim \mathcal{CW}_m(n, \vec I_m)$, such that $ n \geq m$ $\vec{SRS}^H \sim \mathcal{CW}_m(n, \vec I_m)$ for any $m \times m$ unitary matrix $\vec S$. Let $\vec V = \vec{SRS}^H$  and $\vec S$ be a unitary matrix such that, $\vec S^H = (\vec S_1^H, (\vec c^H \vec c)^{-\frac{1}{2}}\vec c)$ where $\vec S_1$ is  $m \times m-1$ matrix and $\vec c \in \mathcal{C}^m$. From \cite{nagarexp}, we obtain $\vec c^H \vec R^{-1} \vec c =(\vec c^H \vec c) v^{mm}$, where the elements of $\vec V^{-1} = (v^{jk})$. Letting $\vec V = \vec {LL}^H$, where $\vec L=(l_{ij})$ is a complex lower triangular matrix with positive diagonal elements, we get, $v^{mm}= l_{mm}^{-2}$.
 Using this result for the case $m=N_R$, 
\begin{equation}
\eta= \frac{E_D}{E_I}\vec c^H \vec R^{-1} \vec c =  \frac{E_D}{E_I}(\vec c^H \vec c) l_{N_R N_R}^{-2}= \frac{E_D}{E_I} \frac{y}{x},
\end{equation}
where $y=\vec c^H \vec c$ and $x^{-1}=l_{N_R N_R}^{-2}$ and are independent.The pdf of $\eta=\frac{(\vec c^H \vec c)}{ l_{N_R N_R}^{2}}\frac{E_D}{E_I}=\frac{y}{x}\frac{E_D}{E_I}$ is can now be obtained by solving the integral using using (\ref{prop}) as,
\ifCLASSOPTIONtwocolumn
\begin{align}\label{pdf2}
\nonumber
f(\eta) =
  & \frac{\frac{E_I}{E_D} ^{N_R \mu} e^{-N_R \kappa \mu}}{\Gamma(N_R \mu) \Gamma(a)}( \mu(1+\kappa))^{N_R \mu} \eta^{N_R \mu -1}  \Gamma(N_R \mu +a)\\
  \nonumber
  & \quad \times ( \mu (1+\kappa) \frac{E_I}{E_D} \eta +1)^{-N_R\mu -a}\\
  & \qquad \times  \,_1F_1(N_R \mu +a, N_R \mu, \frac{N_R \mu^2 \kappa(1+\kappa) \frac{E_I}{E_D} \eta}{ \mu (1+\kappa) \frac{E_I}{E_D} \eta +1} ).
\end{align}
\else
\begin{align}\label{pdf2}
\nonumber
f(\eta) = 
  & \frac{\frac{E_I}{E_D} ^{N_R \mu} e^{-N_R \kappa \mu}}{\Gamma(N_R \mu) \Gamma(a)}( \mu(1+\kappa))^{N_R \mu} \eta^{N_R \mu -1}  \Gamma(N_R \mu +a)\\
  & \quad \times ( \mu (1+\kappa) \frac{E_I}{E_D} \eta +1)^{-N_R\mu -a} \,_1F_1(N_R \mu +a, N_R \mu, \frac{N_R \mu^2 \kappa(1+\kappa) \frac{E_I}{E_D} \eta}{ \mu (1+\kappa) \frac{E_I}{E_D} \eta +1} ).
\end{align}
\fi
Though there are existing results which derive the CDF of this ratio \cite{geetha_outage, ermolova_outage}, obtaining the pdf of SIR $\eta$ by differentiating these CDF expressions is not straightforward. The outage probability for our case can be obtained by substituting $\beta= a$, $\lambda=1$, $\mu=N_R \mu$, $r^a= \frac{E_I}{E_D}$ and $T= \frac{T}{N_R}$ in \cite[Eq.11]{geetha_outage}, as,
\ifCLASSOPTIONtwocolumn
\begin{align}\label{outkmu}
\nonumber
P( \eta <T) &= 
A \Psi_1( a+ N_R \mu, 1; a+1, N_R \mu;  \\
 & \quad \frac{1}{ \mu(1+\kappa)T \frac{E_I}{E_D}}, \frac{N_R \mu^2 (1+\kappa)T \frac{E_I}{E_D}}{ \mu(1+\kappa)T \frac{E_I}{E_D}}),
\end{align}
\else
\begin{align}\label{outkmu}
P( \eta <T) &= 
A \Psi_1( a+ N_R \mu, 1; a+1, N_R \mu; \frac{1}{ \mu(1+\kappa)T \frac{E_I}{E_D}}, \frac{N_R \mu^2 (1+\kappa)T \frac{E_I}{E_D}}{ \mu(1+\kappa)T \frac{E_I}{E_D}}),
\end{align}
\fi
where $A= 1- e^{-N_R \kappa \mu} \frac{\Gamma(N_R \mu +a) }{\Gamma( a+1)\Gamma(N_R \mu)} \frac{(\mu(1+\kappa)T \frac{E_I}{E_D})^{N_R \mu}}{(1+\mu(1+\kappa)T \frac{E_I}{E_D})^{a +N_R \mu}}$ and $\Psi_1$ is the confluent Appell function \cite{geetha_outage}.\\
We apply the following identities from \cite{abr, for1} sequentially to simplify (\ref{pdf2}): a) $\,_1F_1(p+a,p,z)=e^z\,_1F_1(-a, p, -z)$,  b) $\,_1F_1(-a,p, -z)= \frac{a!}{(p)_{a}} L_{a}^{p-1}(-z)$, c) $L_{a}^{p-1}(-z)=\frac{\Gamma(a+p)}{a!}\sum_{k=0}^{a} \frac{(-a)_k (-z)^k}{\Gamma(k+p) k!}$ and d) $(-a)_k =\frac{a!}{(a-k)!}k!$ for $p \in \mathbb{R}^+$ and $a \in \mathbb{Z}^+$. Note, in our case $p=N_R\mu$ and $a \in \mathbb{Z}^+$. Finally, using the Taylor series expansion for the exponential term and the identity for Laplace expansion $M_{\eta}(s)= \int_{0}^{\infty}e^{-s \eta} f(\eta) d(\eta)$, we obtain 
\begin{align*}
M_{\eta}(s)=& \frac{ e^{-N_R \kappa \mu} \Gamma(N_R \mu +a)}{\Gamma(N_R \mu) \Gamma(a)}\sum_{k=0}^{a} \frac{a! (N_R \mu \kappa)^{k}}{(a-k)! k! (N_R \mu)_k} \\ 
  & \, \times \int_{0}^{\infty}e^{-s \eta} \sum_{l=0}^{\infty}  \frac{( \frac{E_I}{E_D} \mu(1+\kappa))^{N_R \mu+k+l} (N_R \mu \kappa)^{l}}{l!}\\
  & \quad \times  \eta^{N_R \mu+k+l -1} ( 1+  \frac{\mu (1+\kappa) E_I \eta}{E_D} )^{-N_R\mu -a-l-k}d \eta.
\end{align*}
The summation and integration can be interchanged by the direct application of Tonelli's theorem \cite{tao}, since $f_l(\eta)=( \frac{E_I}{E_D} \mu(1+\kappa))^{N_R \mu+k+l} \frac{(N_R \mu \kappa)^{k+l}}{l!}   \eta^{N_R \mu+k+l -1} ( \mu (1+\kappa) \frac{E_I}{E_D} \eta +1)^{-N_R\mu -a-l-k} >0 $.
We now use the integration identity $\int_{0}^{\infty}e^{-px}x^{q-1}(1+ax)^{-v}dx= a^{-q}\Gamma(q)U(q,q+1-v, p/a)$ from \cite{int}, to obtain the Laplace expansion of $\eta$ as,
\ifCLASSOPTIONtwocolumn
\begin{align}
\nonumber
M_{\eta}(s)  =& \frac{ e^{-N_R \kappa \mu} \Gamma(N_R \mu +a)}{\Gamma(N_R \mu) \Gamma(a)}\sum_{k=0}^{a} \frac{a! (N_R \mu \kappa)^{k}}{(a-k)! k! (N_R \mu)_k} \\  
  & \quad \times \sum_{l=0}^{\infty} \frac{(N_R \mu \kappa)^{l} \Gamma(b)}{l!} U\left(b , 1-a,\frac{s}{\mu (1+\kappa) \frac{E_I}{E_D}}\right), \label{le}
\end{align}
\else
\begin{align}
\nonumber
M_{\eta}(s)= & \frac{ e^{-N_R \kappa \mu} \Gamma(N_R \mu +a)}{\Gamma(N_R \mu) \Gamma(a)}\sum_{k=0}^{a} \frac{a!}{(a-k)! k! (N_R \mu)_k} \sum_{l=0}^{\infty} (\frac{E_I}{E_D}  \mu(1+\kappa))^{b}\\ 
\nonumber
 &  \frac{(N_R \mu \kappa)^{k+l}}{l!} (\frac{E_I}{E_D}  \mu(1+\kappa))^{-b}  \Gamma(b)U(b , b  +1 -b-a,\frac{s}{\mu (1+\kappa) \frac{E_I}{E_D}})\\
   \nonumber
  =& \frac{ e^{-N_R \kappa \mu} \Gamma(N_R \mu +a)}{\Gamma(N_R \mu) \Gamma(a)}\sum_{k=0}^{a} \frac{a!}{(a-k)! k! (N_R \mu)_k}  \sum_{l=0}^{\infty} \frac{(N_R \mu \kappa)^{k+l} \Gamma(b)}{l!}U(b , 1-a,\frac{s}{\mu (1+\kappa) \frac{E_I}{E_D}}). \label{le}
\end{align}
\fi
where $b=N_R \mu +k +l$. 
Substituting the expression for the Laplace expansion from (\ref{le}) in (\ref{pe}), we obtain the average SER as
\begin{align}
\nonumber
SER&=\sum_{m=1}^{5} a_m   \frac{ e^{-N_R \kappa \mu} \Gamma(N_R \mu +a)}{\Gamma(N_R \mu) \Gamma(a)}\sum_{k=0}^{a} \frac{a!}{(a-k)! k! (N_R \mu)_k} \\  
  & \quad \times \sum_{l=0}^{\infty} \frac{(N_R \mu \kappa)^{k+l}\Gamma(b)}{l!} U\left(b, 1-a,\frac{b_m}{\mu (1+\kappa) \frac{E_I}{E_D}}\right),\label{ser}
\end{align}
where $U(.)$ is the Tricomi Hypergeometric function \cite{hanzo}. 
Since $U(a+1,c+1,x) < \frac{-1}{c}U(a,c,x)$ for $a>0>c$ and $x >0$ from \cite{baricz}, we obtain 
\begin{align*}
&U(b, 1-a,\frac{b_m}{\mu (1+\kappa) \frac{E_I}{E_D}})\\ & <\frac{U(N_R \mu +k,1-a-l, \frac{b_m}{\mu (1+\kappa) \frac{E_I}{E_D}})}{(N_I-N_R)_l}.
\end{align*}
Also, since $U(m, n-k,x)$ decreases monotonically with $k$ \cite{hanzo}, $\sum_{l=0}^{\infty} \frac{(N_R \mu \kappa)^{k+l}}{l!} \Gamma(b)U(b, 1-a,\frac{b_m}{\mu (1+\kappa)\frac{E_I}{E_D}}) <  \sum_{l=0}^{\infty} \frac{(N_R \mu \kappa)^{k+l}}{l!} \Gamma(b)\frac{U(N_R \mu +k,N_R -N_I, \frac{b_m}{\mu (1+\kappa) \frac{E_I}{E_D}})}{(N_I-N_R)_l}$ which is a converging series. Hence the infinite summation can be truncated to a finite series with arbitrarily low truncation error. 

\section{Numerical Results}
The derived SER expressions are verified using Monte-Carlo simulations. The total interference power is denoted as $E_I''$, from which average interference power per interferer is obtained as $E_I'=E_I''/N_I$. The mean energy of the received signal, $E_D$, is taken to be unity without loss of generality. SIR is given by $E_D/E_I''$. The determinant of $\mathbf N$ matrix whose entries are given by (\ref{Nfin}) is determined with the infinite summation truncated to $T_1=T_2=70$. For a SER of $10^{-5}$, with $T_1=100$ and $T_2=100$ terms, the numerical evaluation completes in $40$ seconds\footnote{in MATLAB R2015b run in an iMac with 2.8GHz with intel i5 core and 8 GB RAM in Sierra OS}. Also, for Rayleigh faded interferers, the expressions have only a single infinite series, which takes a maximum of $2$ seconds for evaluation. On the other hand, Monte-Carlo simulations take close to 500 seconds. By substituting the determinant of $\mathbf N$ matrix in (\ref{MGFfinal}), the Laplace transform is evaluated for $s=b_l$, $\forall l=1 \; to \; 5$. These Laplace transform values are substituted in (\ref{pe}) and theoretical SER is calculated for values of signal to noise ratio (SNR) in the range $5$ dB to $25$  dB. For the Monte-Carlo simulation, the deterministic matrix $\mathbf{M}'$ is first obtained with unit magnitude and uniform phase, satisfying the condition $\text{tr}(\mathbf{M}'\mathbf{M}'^H)=N_RN_I$. This matrix is fixed during a set of simulations.
\par
\begin{figure}
\centering
\subcaptionbox{$E_I''=-1dB, \kappa_s=3$ \label{fig1}}{\includegraphics[height=1.8in, width=3in]{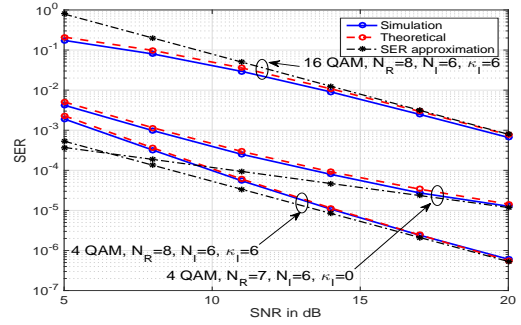}}
\hfill
\subcaptionbox{ $ E_I''=-1dB, \kappa_s=3$  \label{fig2}}{\includegraphics[height=1.8in, width=3in]{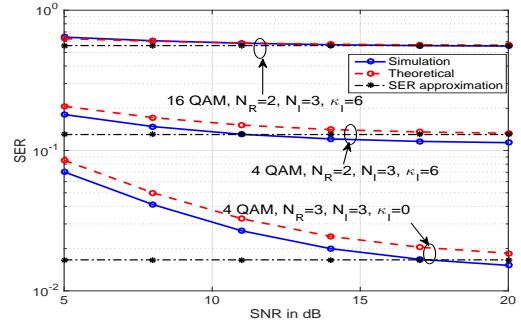}}
\caption{SER  for $N_R \geq N_I$ and $N_R \leq N_I$}
\label{FIG1}
\end{figure}

 For the case of equal power and uncorrelated interferers, a close match between the theoretical and simulated SER is observed in Fig. \ref{FIG1}(\subref{fig1}) and Fig. \ref{FIG1}(\subref{fig2}). The exponential approximation $\textstyle Q(x) \approx  \textstyle \frac{1}{12} e^{-\frac{1}{2}x^2} +\frac{1}{4} e^{-\frac{2}{3}x^2}$, from  \cite{bounds} provides a very tight upper bound for values of $x >0.5$ and the bound becomes tighter as $x$ increases. Since, for $N_I > N_R$, the average SINR is much lower when compared to the case $N_R > N_I$, we can observe a small mismatch between the theoretical and the simulated SER in Fig. \ref{FIG1}(\subref{fig2}).  Also, the SER approximation plot in  Fig. \ref{FIG1} is tight beyond $15$ dB. For Rayleigh interferers, i.e., $\kappa_I=0$, the SER approximation computed using (\ref{highSNR2}), match with the simulation results at high SNR, for $N_R > N_I$ as seen from Fig. \ref{FIG1}(\subref{fig1}) and Fig. \ref{FIG1}(\subref{fig1}). We can also see from Fig. \ref{FIG1} that, when the interference power dominates the noise power, as is the case when $E_I''=-1$ dB, the SER approximation is tight even at $15$ dB SNR. 
\par
\begin{figure}
\centering
\subcaptionbox{ correlated Rayleigh interferers $ \kappa_s=3, N_R=4, N_I=2, E_I''=  -1$ dB, $\rho=0.6$  \label{fig3}}{\includegraphics[height=1.8in, width=3in]{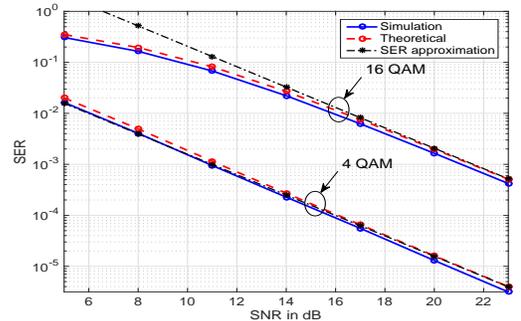}}
\caption{SER for correlated Rayleigh interferers}
\label{FIG2}
\end{figure}

\begin{figure}
\centering
\subcaptionbox{ mixture of two Rician and two Rayleigh interferers $ \kappa_s=2, \kappa_i=5,N_R=5, N_I=4, E_I''=  -1, -1,5, -2, -2.5$ dB \label{fig4}}{\includegraphics[height=1.8in, width=3in]{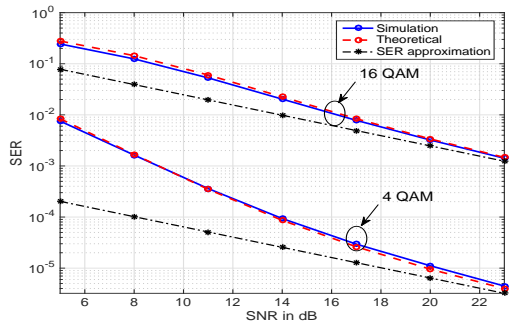}}
\hfill
\subcaptionbox{ unequal power Rayleigh interferers $ \kappa_s=2, N_R=3, N_I=5, E_I''= -6, -9, -10, -10, -10$ dB \label{fig5}}{\includegraphics[height=1.8in, width=3in]{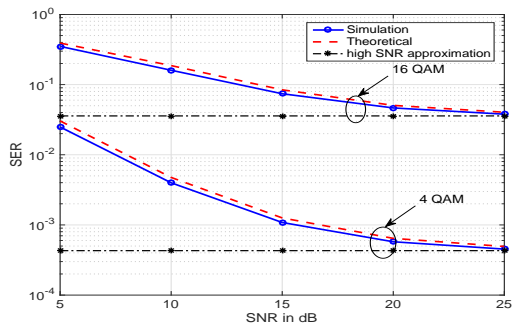}}
\caption{SER for unequal power Rayleigh/Rician interferers}
\label{FIG3}
\end{figure}

 Similar Monte-Carlo simulations are performed for the case of correlated and/or unequal power interferers. The difference is that, random covariance matrix $\mathbf{R}$ of the interference terms plus the noise term is now calculated using (\ref{R1}). Also note that the determinant of $\mathbf N$ matrix whose entries are given by (\ref{Nunequal}).  From Fig. \ref{FIG2}(\subref{fig3}), we can observe that for correlated Rayleigh faded interferers, SER computed by means of (\ref{MGFfinal1}) and (\ref{Nunequal}) matches the simulated values. We can also see that the approximation for $\sigma^2 \approx 0$ computed using (\ref{highSNRunequal}) gives good match with the simulated results for higher values of SNR. We consider exponential correlation between interferers \cite{sri}, i.e., $\vec \Psi(i,j)=\rho^{|i-j|}$ and  $0 \leq \rho \leq 1$, where $\vec \Psi$ is the interferer covariance matrix. Similar results can be obtained for unequal power interferers and is not give here due to space constraints. In Fig. \ref{FIG3}(\subref{fig4}), we studied the case of mix of Rayleigh and Rician faded unequal power interferers.  Here also, the SER approximation given by (\ref{highSNRunequal}) gives a good match to the simulated SER for high SNR values.  For Rayleigh faded interferers with unequal power and $N_I > N_R$, the approximation in Section IV.B gives a fairly good match to the theoretical values as seen from Fig. \ref{FIG3}(\subref{fig5}), as long as the interferer powers do not vary widely.
\par It is known that OC trades off the effect of noise and interference at the receiver. In the absence of noise it maximizes the average SIR and specializes to a ZF receiver, which is the ideal receiver for interference cancellation. In the presence of noise, OC balances between noise cancellation and interference cancellation. This results in a sub-optimal average SIR. A better SIR translates to a lower SER and hence we observe a lower SER for both $4$-QAM and $16$ QAM, for $N_I >N_R$ and $\sigma^2=0$. This is captured by the fact that SER approximation that assumes $\sigma^2=0$ forms a lower bound for the SER values of $N_I > N_R$ with non-zero $\sigma^2$. This can be observed in Fig. \ref{FIG1}(\subref{fig2}). Also, note that the bound becomes tighter for large values of SNR. In this regime, OC mimics the performance of ZF. In the case of $N_R > N_I$, the approximation is ad-hoc and computationally less intensive due to the absence of determinant evaluation.
\par
\begin{figure}
\centering
\subcaptionbox{ $E_I''=-1$ dB, $\kappa_s=6, \kappa_I=3$ \label{fig6}}{\includegraphics[height=1.8in, width=3in]{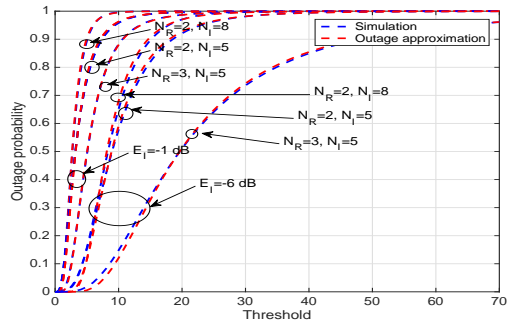}}
\hfill
\subcaptionbox{ $N_R=2, N_I=5, E_I''=-1$ dB, $\kappa_s=6$  \label{fig7}}{\includegraphics[height=1.8in, width=3in]{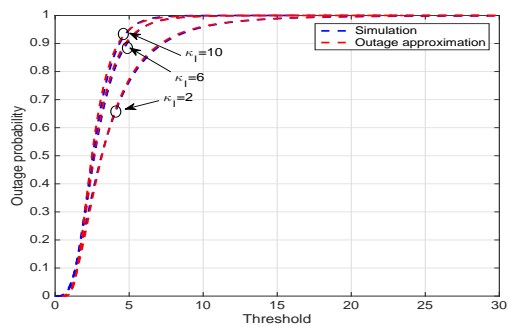}}
\caption{Outage probability for Rician interferers}
\label{FIG4}
\end{figure}

\begin{figure}
\centering
 \includegraphics[height=1.8in, width=3in]{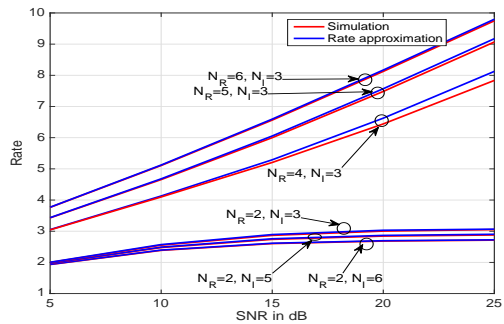}
 \caption{ Rate vs SNR, $ \kappa_s=8, \kappa_I=3, E_I=-4$ dB}
\label{FIG5}
\end{figure}

\begin{figure}
\includegraphics[width=3.5in, height=1.8in]{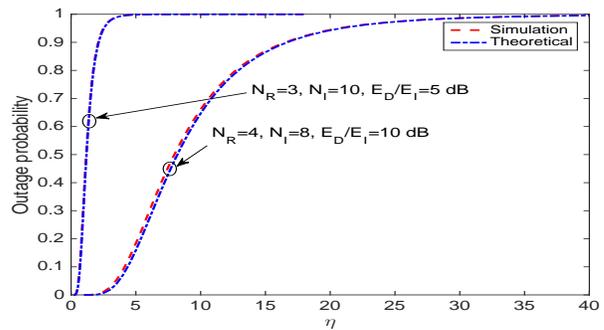}
\caption{Outage probability for $\kappa-\mu$ faded users $ \kappa=1.5, \mu=4.0$}
\label{FIG6}
\end{figure}

\begin{figure}[!h]
\centering
\includegraphics[width=3.5in, height=1.8in]{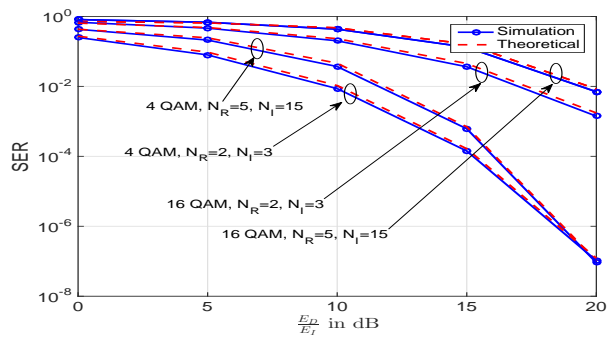}
\caption{SER vs $\frac{E_D}{E_I}$ for $ \kappa=4.0, \mu=5.3$}
\label{FIG7}
\end{figure}

 The variation of the outage probability is shown in Fig. \ref{FIG4}(\subref{fig6}) for changes in $N_R$ and $N_I$. As discussed in I3, the outage probability decreases with an increase in $N_R$ or a decrease in $N_I$. Also, as discussed in I1, the outage probability decreases with a decrease in $E_I$. In Fig. \ref{FIG4}(\subref{fig7}), $\kappa_I$, the Rician parameter of the interferers is varied. We can observe that the outage probability decreases with a decrease in $\kappa_I$. Finally, the Monte-Carlo simulations and approximate expressions for rate are shown in Fig. \ref{FIG5}. We can observe that the rate increases with an increase in $N_R$ or SNR or a decrease in $N_I$. Finally, from Fig. \ref{FIG6} and \ref{FIG7}, we can observe that the OP expressions (\ref{outkmu}) and SER expressions (\ref{ser}) for $\kappa-\mu$ fading users match the simulations.

\section{Applications}
Device-to-device (D$2$D) communication enables a pair of closely located mobile users to establish a direct link for their user-plane traffic without going through the entire network infrastructure, while reusing the spectrum allocated for traditional cellular communication. Due to the existence of a LoS component, it is appropriate to model D$2$D links by Rician fading. In fact, in \cite{ wang2018joint} and references therein, several performance metrics were analyzed in Rician fading D$2$D systems. In D$2$D underlying cellular networks, strong interference may occur between cellular and D$2$D links sharing the same spectrum. To mitigate the interference, many resource allocation schemes have been proposed in \cite{wang2017joint,wang2018joint, liang2017d2d}. Similar to \cite{wang2018joint}, we consider a D$2$D-enabled communications network, where M single-antenna cellular users (CUE) perform high-capacity uplink cellular communications with the base station (BS) with $N_R$ antennas. In \cite{wang2017joint,liang2017d2d,wang2018joint}, only one antenna at BS is considered.  But it makes practical sense to consider the BS to be equipped with more than one antenna. Also, consider K pairs of single-antenna users doing local data exchange in the form of D$2$D communications. These are denoted as DUEs or D$2$D users. The K DUEs can reuse the M sub-channels in the underlay mode. Since the BS is equipped with more than one antenna, the interference at the BS can be cancelled by means of OC.
\par Let $\vec h_{m,B}$ denote the $N_R \times 1$ Rician channel (with Rician factor $\kappa_{m,B}$) from the $m^{th}$ CUE to the BS. $h_k$ denotes the channel fast fading coefficient between the $k$th DUE pair. Due to the presence of strong LoS component between the D$2$D receiver (Rx) and transmitter (Tx), $h_k$ is assumed to be Rician with Rician factor $\kappa$ as in \cite{wang2018joint}. $\vec h_{k, B}$ denote the $N_R \times 1$ interferer channel fast fading coefficients from the $k$th DUE Tx to BS. $h_{m,k}$ denotes the channel fast fading coefficient (with Rician factor $\kappa_{m,k}$) from the $m$th CUE to the the Rx of the $k$th DUE. This alone is assumed to be Rayleigh faded as in \cite{wang2018joint}. Also, let $s_{m,B}$, $s_k$, $ s_{k,B}$ and $s_{m,k}$ denote the corresponding slow fading coefficients. Since we consider a high-speed scenario where fast-fadings are difficult to estimate and only slow-fadings are available, we can assume the slow-fadings to be known constants and the fast-fadings to be random variables. Denote the CUE set as $\mathcal M =\{1,..., M\}$ and the DUE set as $\mathcal K =\{1,...,K\}$. The SINR at the output of OC at the BS is
\begin{align}
\nonumber
    \gamma_m^c &= P_m^c s_{m,B}\vec h_{m,B}^H(\underset{k \in \mathcal{K}}{\textstyle\sum} \rho_{m,k}P_k^d s_{k,B} \vec h_{k, B} \vec h_{k, B} ^H +\sigma^2 \vec I)^{-1} \vec h_{m,B}
\end{align}
and the SINR at the the $k$th D$2$D Rx is
\begin{align}
    \gamma_k^d= \frac{P_k^d s_k|h_{k}|^2}{\sum_{m \in \mathcal{M}}\rho_{m,k} P_m^c s_{m,k}|h_{m,k}|^2 +\sigma^2 },
\end{align}
where $P_m^c$ and $P^d_k$ denote transmit powers of the $m$th CUE and the $k$th DUE, respectively, $\sigma^2$ is the noise power and $\rho_{m,k}$ is the spectrum allocation indicator with $\rho_{m,k}=1$ indicating the $k$th DUE reuses the spectrum of the $m$th CUE and $\rho_{m,k}=0$ otherwise. The ergodic capacity of the $m$th CUE is given by $C_m = \mathbb E[ \text{log}_2 (1 + \gamma_m^c)]$.  We maximize the sum ergodic capacity of $M$ CUEs while guaranteeing the minimum reliability for each DUE. In addition, we set a minimum capacity requirement for each CUE. The reliability of DUEs is guaranteed through controlling the probability of outage events, where its received SINR $\gamma_k^d$ is below a predetermined threshold $\gamma_0^d$ .
Hence the radio resource allocation problem in vehicular networks is formulated as \cite{wang2017joint, wang2018joint}
\begin{align}
\nonumber
   & \underset{\rho_{m,k} P_m^c, P_k^d}{\text{max}} \sum_{m \in \mathcal{M}} \mathbb E[\text{log}_2 (1+\gamma_m^c)],\\
   \nonumber
   & \text{s.t.} \: \mathbb E[\text{log}_2 (1+\gamma_m^c)] \geq r_0^c, \: \forall m \in \mathcal{M},\\
\nonumber
   & Pr\{\gamma_k^d \leq  \gamma_0^d\} \leq  p_0, \forall k \in \mathcal{K},\\
\nonumber
   & 0 \leq P_m^c \leq P_{max}^c, \forall m \in \mathcal{M} \: \text{and} \:
   0 \leq P_k^d \leq P_{max}^d, \forall k \in \mathcal{K},\\
\nonumber
   &\sum_{\forall m \in \mathcal{M}} \rho_{m,k} \leq 1, \: \rho_{m,k} \in \{0,1\}, \: \forall m \in \mathcal{M},\\
   &\sum_{\forall k \in \mathcal{K}} \rho_{m,k} \leq 1, \: \rho_{m,k} \in \{0,1\}, \: \forall k \in \mathcal{K},
\end{align}
 where $r^c$ is the minimum capacity requirement of the rate intensive CUEs and $\gamma_0^d$ is the minimum SINR needed by the DUEs to establish a reliable link. $p_0$ is the tolerable outage probability of the D$2$D links. $P_{max}^c$ and  $P_{max}^d$ are the maximum transmit powers of the CUE and DUE, respectively. The last two constraints mathematically model the assumption that the spectrum of one CUE can only be shared with a single DUE and one DUE is only allowed to access the spectrum of a single CUE.
\par  The power allocation problem can be written for $k$th DUE sharing the band of the $m$th CUE as
 \begin{align}
 \nonumber
    & \underset{ P_m^c, P_k^d}{\text{max}}\mathbb E[\text{log}_2 (1+\gamma_m^c)],\\
\nonumber
   & \text{s.t.}\: Pr\{\gamma_k^d \leq  \gamma_0^d\} \leq  p_0,\\
   & 0 \leq P_m^c \leq P_{max}^c \: \text{and} \:0 \leq P_k^d \leq P_{max}^d. 
 \end{align}
Since the D$2$D links are Rician and the link between the the CUE and Rx of the D$2$D link is Rayleigh as in \cite{wang2018joint}, the reliability constraint of the $k$th DUE, $Pr\{\gamma_k^d \leq  \gamma_0^d\} \leq  p_0$, is given by \cite[Eq 6]{wang2018joint}
    $P(\gamma_k^d \leq  \gamma_0^d) = \frac{\theta}{\theta+1}e^{-\frac{\kappa}{\theta+1}} =f(\theta)$,
where $\theta=\frac{\gamma_0^dP_m^c s_{m,k}}{P_k^d s_k}$. From \cite[Eq.9]{wang2018joint}, we can obtain an upper bound of $P_m^c$ as
$P_m^c \leq \frac{s_k\theta_0}{s_{m,k}\gamma_0^d}P_k^d$,
where $\theta_0=f^{-1}(p_0)$ and $\beta=\frac{s_k\theta_0}{s{m,k}\gamma_0^d}$.
Using the same reasoning as in \cite{wang2018joint}, the optimal power allocation is  obtained as
\begin{align}
    (P_{m}^{c^*}, P_{k}^{d^*}) &= \begin{cases}
    (P_{max}^c,\frac{P_{max}^c}{\beta}), \: \text{if} \, \beta > \frac{P_{max}^c}{P_{max}^d},\\
    (P_{max}^d \beta, P_{max}^d), \: \text{if} \, \beta < \frac{P_{max}^c}{P_{max}^d}.
    \end{cases}
\end{align}
With the optimal power allocations
$(P_{m}^{c^*}, P_{k}^{d^*})$, under each given channel allocation, we aim to maximize the cellular rates by searching over all possible channel allocation schemes. We first determine the ergodic capacity of $m$th CUE when it shares spectrum with $k$th DUE is $C_{m,k}$ for a power allocation $(P_{m}^{c^*}, P_{k}^{d^*})$. Since, OC is used at the BS, the rate $ C_{m,k}(P_{m}^{c^*}, P_{k}^{d^*})= \mathbb E[\text{log}_2 (1+\gamma_m^c{}^*)]$, where $\gamma_m^c{}^*$ is the SINR at the output of OC for the power allocation $(P_{m}^{c^*}, P_{k}^{d^*})$. This rate can be easily determined by means of the rate expressions (\ref{rate}). Without the rate expressions, it wouldn't have been possible to consider an interference cancellation scheme like OC at the BS. To guarantee the minimum transmission rates for CUEs and DUEs, we have to exclude those CUE-DUE pairs unable to meet the rate requirements even with the optimal transmitting powers. This condition is interpreted as $\rho_{m,k} =0$ if the pairing of CUE $m$ with DUE $k$ violates the rate constraints. If for one such pair $(m, k)$, $ C_{m,k}(P_{m}^{c^*}, P_{k}^{d^*})$ is lesser than the minimum rate  $r_0^c$, then the rate is replaced by $-\infty$. The exact procedure is as discussed in \cite{liang2017d2d, wang2017joint}. Hence, 
\begin{equation}
    C_{m,k}^{*}= \begin{cases}
         C_{m,k}(P_{m}^{c^*}, P_{k}^{d^*}), \: \text{if} \: C_{m,k}(P_{m}^{c^*}, P_{k}^{d^*}) \geq r_0^c,\\
        -\infty, \: \text{otherwise}
    \end{cases}
\end{equation}
The spectrum allocation problem becomes
\begin{align}
\nonumber
   & \underset{\rho_{m,k}}{\text{max}} \sum_{m \in \mathcal{M}}C_{m,k}^{*},\\
\nonumber
   &\sum_{\forall m \in \mathcal{M}} \rho_{m,k} \leq 1, \: \rho_{m,k} \in \{0,1\}, \: \forall m \in \mathcal{M},\\
   \nonumber
   &\sum_{\forall k \in \mathcal{K}} \rho_{m,k} \leq 1, \: \rho_{m,k} \in \{0,1\}, \: \forall k \in \mathcal{K}.
\end{align}
This is a maximum weight bipartite matching problem and can be solved by the Hungarian method as in \cite{liang2017d2d, wang2017joint, wang2018joint}. Simulation results are presented to evaluate the performance of joint power and channel allocation scheme. The simulation parameters are as follows: $M = 40$, $N = 20$, $r_0^c = 0.1$ bps/Hz,  $\gamma_0 = 5$dB, $p_0 = 0.1$, $P^c_{max}= P^d_{max}=25$ dBm and $\kappa_{m,B}=\kappa_{k,B}=0.1$. BS antenna gain is $8$dBi and the user antenna gain is $3$ dBi. The  noise power is $\sigma^2=-114$ dBm. The loss exponents of the links to the BS and user antenna are $3.76$ and $2.27$, respectively and the shadow fading standard deviations are $8$ dB and $3$dB, respectively. 
\begin{table}[h]
\begin{tabular}{|l|l|l|}
\hline
$\kappa$ & Sum rate of CUE for $N_R=3$ & Sum rate of CUE for $N_R=2$  \\
\hline
$0$ & $815.6465$ & $780.0624$ \\
$0.5$ & $816.7338$ & $781.1835$ \\
$1$ & $818.1748$ & $782.7159$ \\
$2$ & $819.3933$ & $783.8709$ \\
$3$ & $819.4748$ & $783.9478$\\
\hline
\end{tabular}
\caption{Sum rate of CUE in bps/Hz} 
\label{table1}
\end{table}
It can be observed from the Table \ref{table1} that the sum rate in bps/Hz increases with increase in $N_R$ and the Rician factor $\kappa$ between the D$2$D pairs. Also, the utility of our expressions lies in the fact that it can be used in any application, where the user and interferers undergo Rician fading and OC is employed such as this application and Vehicular Ad-hoc Network (VANET).
\section{Conclusions}
Approximate SER, outage probability and rate expressions have been derived for OC for the case of Rician faded users and a) Rician faded interferers, b) mixture of Rician and Rayleigh faded interferers and c) Rayleigh faded interferers when the interferers are correlated/uncorrelated and have equal or unequal powers. SER is also derived for an interference-limited scenario and the expressions obtained are significantly simpler than the existing expressions. The Monte-Carlo simulation closely match the derived results. We believe extending this analysis to take into account receiver side correlation may be interesting future work. An application where our results have significant utility is also discussed.

\appendices
\section{Simplification of $N_{i,j}$}\label{Nsimp}
We first substitute the value of $h(t,x)$ from (\ref{ht}) for $N_{ij}$ $j=1$ and then use the following identities \cite{int}, for $p<q$  and $\text{Re}(s)>0$,
\begin{multline} \label{prop}
 \int_{0}^{\infty}e^{-x}x^{s-1}\,_pF_q(a_1,..,a_p;b_1,..,b_q;ax)dx\\
 =\Gamma(s)\,_{p+1}F_q(s,a_1,..,a_p;b_1,..,b_q;a),
\end{multline}
\begin{align} \label{prop1}
 \int_{0}^{\infty}e^{-x}x^{s-1}dx =\Gamma(s),
\end{align}
to solve the integrals in $N_{i,j}$ entries for $ j=2,..,n_1$ in (\ref{MGFfinal}). For $j=1$, $i=1,...,L$ and $i=L+1,...,n_1$, the integrals to be solved  are of the form,
\begin{equation}\label{I2}
 I=\int_{0}^{\infty}\frac{p+x}{v-x}e^{-x}x^z(_0F_1(q,w_ix))(_1F_1(1;N_R;\frac{u}{x-v}))dx,
\end{equation}
where $p=\sigma^2/E_I$, $q=n_2-n_1+1$, $u=aN_RsE_D/E_I$, $v=bsE_D/E_I-\sigma^2/E_I$ and $z$ is a positive integer greater than zero.
To obtain a solution for $I$, we substitute the series expansion for $_0F_1$ and $_1F_1$, and interchange summations and integration. The integral to be solved becomes,
\begin{equation}\label{Iint}
 I=\sum_{k=0}^{\infty}\frac{w_i^k}{(q)_kk!}\bigg[\sum_{l=0}^{\infty}\frac{u^l(1)_l}{(N_R)_ll!}\int_{0}^{\infty}\frac{p+x}{v-x}e^{-x}\frac{x^{k+z}}{(x-v)^l}dx\bigg].
\end{equation}
The justification for the interchange of summations and integration can be done in two steps. We first expand the $\,_0F_1$ term and apply Tonelli's theorem to justify the exchange of the first summation. The $\,_1F_1$ term is now expanded and the summation is interchanged with the integration using Lebesgue dominated convergence theorem.
Let $ A1=\int_{0}^{\infty}\frac{p}{v-x}e^{-x}\frac{x^{k+z}}{(x-v)^l}dx$ and
$ A2=\int_{0}^{\infty}\frac{x}{v-x}e^{-x}\frac{x^{k+z}}{(x-v)^l}dx$.
The Tricomi function or confluent Hyper-geometric function of the second kind is given by \cite{int},  $U(\alpha, \gamma, z)=\frac{1}{\Gamma(\alpha)}\int_{0}^{\infty}e^{-zt}t^{\alpha-1}(1+t)^{\gamma-\alpha-1}dt$ for $\text{Re} (\alpha) > 0, \text{Re} (z) >0.$
In our case $-v >0$ and $k+z>0$. Hence using the above identity, $A_1$ and $A_2$ can be simplified as,
 \begin{equation}\label{A1int}
 A1=-p\Gamma(k+z+1)(-v)^{l-k-z}U(k+z+1,k+z+1-l,-v),
 \end{equation}
 \begin{equation}\label{A2int}
 A2=-\Gamma(k+z+2)(-v)^{l-k-z-1}U(k+z+2,k+z+2-l,-v).
 \end{equation}
 Further, using the functional identity  $U(a,b,z)=z^{1-b}U(a-b+1,2-b,z)$ from \cite{abr}, (\ref{A1int}) and (\ref{A2int}) are simplified and substituted back in (\ref{Iint}) to obtain,
 \begin{multline}\label{fullI}
 I=\sum_{k=0}^{\infty}\frac{w_i^k}{(q)_kk!}\Bigg[\sum_{l=0}^{\infty}\frac{u^l(1)_l}{(N_R)_ll!} \bigg[-\Gamma (k+z+2)\\
 \times  U(l+1,l-k-z,-v)\\ -p \Gamma (k+z+1) U(l+1,-k-z+l+1,-v)\bigg]\Bigg].
\end{multline}
To reduce the computation time, we can use the recurrence identity for Tricomi hypergeometric functions given in \cite{for}. To prove the convergence of the above infinite summation, first consider the summation $I_1= \sum_{k=0}^{\infty}\frac{w_i^k}{(q)_kk!}\Bigg[\sum_{l=0}^{\infty}\frac{|u|^l(1)_l}{(N_R)_ll!} \Gamma (k+z+2) U(l+1,l-k-z,-v)\Bigg]$. From Theorem 3 in \cite{joshihyper}, we get the identity $ U(a,b,x)< x^{-a}$ for $ x>0, \: a>0$ and $a-b+1>0$.  In our case, we can see that $a=l+1 >0$ and $a-b+1=k+z+2>0$ and $x=-v >0$. 
Therefore, \begin{align*}
I_1 &< \sum_{k=0}^{\infty}\frac{w_i^k}{(q)_kk!}\sum_{l=0}^{\infty}\frac{|u|^l(1)_l}{(N_R)_ll!} \Gamma (k+z+2)(-v)^{l+1}\\
&=(-v)\Gamma(z+2)\sum_{k=0}^{\infty}\frac{w_i^k (z+2)_k}{(q)_kk!}\sum_{l=0}^{\infty}\frac{|uv|^l(1)_l}{(N_R)_ll!}\\
&=(-v)\Gamma(z+2)\,_1F_1(z+2, q, w_i)\,_1F_1(1, N_R, |uv|).
\end{align*}  
The last equality is obtained from the series expansion definition of $\,_1F_1$ Hypergeometric function \cite{abr}. A similar argument can be used to prove the absolute convergence of the other infinite summation.  
Hence, I is convergent, which implies that we can truncate the double summation to $T_1$ and $T_2$ values such that $I- \sum_{l=0}^{T_1}\sum_{k=0}^{T_2}\frac{u^l}{(N_R)_l}\frac{w_i^k}{(q)_k} \bigg[-(k+1) U(l+1,l-k,-v)  -p  U(l+1,-k+l+1,-v)\bigg] \quad  \leq \epsilon$ for any  $\epsilon >0$. 
Hence, the simplified $N_{i,j}$ entry is given by (\ref{Nfin}). 

\subsection*{Analysis of truncation error}
An exact analysis of the truncation error is mathematically intractable. So, we upper bound the truncation error by an upper bound and determine how the bound varies with various parameters. 
The magnitude of the error in truncation is
\begin{align*}
E(T_1, T_2) &= \Big|\sum_{k=0}^{\infty}\frac{w_i^k}{(q)_kk!}\sum_{l=0}^{\infty}\frac{u^l(1)_l}{(N_R)_ll!} \Gamma (k+z+2)\\ & U(l+1,l-k-z,-v) - \sum_{k=0}^{T_1}\frac{w_i^k}{(q)_kk!}\sum_{l=0}^{T_2}\frac{u^l(1)_l}{(N_R)_ll!}\\ 
&\Gamma (k+z+2) U(l+1,l-k-z,-v)\Big|.
\end{align*}
From Theorem 3 in \cite{joshihyper}, we get the identity $ U(a,b,x)< x^{-a}$ for $ x>0, \: a>0$ and $a-b+1>0$.  In our case, we can see that $a=l+1 >0$ and $a-b+1=k+z+2>0$ and $x=-v >0$. Hence, the truncation error can be upper bounded by
\begin{align*}
E(T_1, T_2) &< |v|\sum_{k=0}^{T_1}\frac{w_i^k}{(q)_kk!}\Bigg[\sum_{l=T_2+1}^{\infty}\frac{|uv|^l(1)_l}{(N_R)_ll!} \Gamma (k+z+2)\Bigg]\\ 
&+ |v| \sum_{k=T_1+1}^{\infty}\frac{w_i^k}{(q)_kk!}\Bigg[\sum_{l=0}^{T_2}\frac{|uv|^l(1)_l}{(N_R)_ll!} \Gamma (k+z+2)\Bigg]\\
&+ |v|\sum_{k=T_1+1}^{\infty}\frac{w_i^k}{(q)_kk!}\Bigg[\sum_{l=T_2+1}^{\infty}\frac{|u|^l(1)_l}{(N_R)_ll!} \Gamma (k+z+2) \Bigg].
\end{align*} 
Combining the last two terms, then upper-bounding the first term, and finally using hypergeometric expansion identity for $\,_1F_1$, we obtain
\begin{align*}
E(T_1, T_2) &< |v|\Gamma(z+2)[\,_1F_1(z+2, q, w_i)\frac{ |uv|^{T_2+1}}{(N_R)_{T_2+1}}\\
&\,_1F_1(1, N_R+T_2+1, |uv|)+  \frac{(z+2)_{T_1+1} w_i^{T_1+1}}{(q)_{T_1+1} (T_1+1)!}\\ & \,_1F_1(1, N_R, |uv|)\\ &\quad  \,_2F_2(z+T_1+3,1, q+T_1+1, T_1+2, w_i)].
\end{align*} 
As the Rician factor $\kappa_i$ at the interferers decreases, the eigenvalues of the centrality matrix $\vec{MM}^H$, given by $w_i$, decreases. Therefore, $E(T_1, T_2)$ decreases and the bound becomes tighter. This implies that we need lesser terms in the infinite summation as $\kappa_i$ increases.
Also, observe that a decrease in $|u|$ or $|v|$ decreases $E(T_1, T_2)$. A decrease in $|u|$ or $|v|$ is true for an increase in $E_I$ or $\sigma^2$.

\section{Approximation for Rayleigh interferers} \label{zero}
Consider the expression to be simplified,
 $M_{\eta}(s)=c|\vec{N}_{\sigma^2=0, L=0}|$,
where
\begin{equation*}
c=\frac{((n_2-n_1)!)^{-n_1}}{\prod_{i=1}^{n_1}(n_1-i)!}(-1)^{N_R}(\sigma^2/E_I)^{(N_R-n_1)}
\end{equation*} 
and $\vec{N}_{\sigma^2=0, L=0}$ is from (\ref{Nfin}) for $L=0$.
First, the common terms inside each column or row of the determinant are taken out of the determinant and canceled with the existing terms in the constant $c$. All columns $j=2,...,n_1$ are flipped and all rows $i=1,..,n_1$ are flipped. The term $\Gamma(n_2-n_1+i+j-1)$ is then removed from each row to obtain $\vec {\tilde N}$. Now,
$ M_{\eta}(s)=c|\vec {\tilde N}|$,
where
\begin{equation}
 c=\frac{\prod_{i=1}^{n_1}(n_2-n_1+i)! \left(\frac{\sigma^2}{E_I}\right)^{(N_R-n_1)}}{\prod_{i=1}^{n_1}(n_1-i)! \prod_{i=1}^{n_1}(n_2-i)!}(-1)^{N_R+1},
\end{equation}
\begin{equation}\label{Ninitial}
\vec {\tilde N}_{i,j}=\begin{cases}
A(i) \;&  j=1,\, i=1,..., n_1,\\
1, \:  &j=2, \, i=1,..., n_1,\\
\prod_{k=1}^{j-2}(n_2-n_1+i+k),  &j=3,...,n_1, 1 \leq i \leq n_1,
         \end{cases}
\end{equation}
where
\ifCLASSOPTIONtwocolumn
$A(i)=\frac{1}{(n_2-n_1+i)!}\Bigg[ \sum_{l=0}^{T_1}\frac{(aN_RsE_D/E_I)^l}{(N_R)_l}\\
\bigg[-\Gamma(n_2-N_R+i+1)U(l+1,l-n_2+N_R-i+1,-bsE_D/E_I)\bigg]\\
	     \quad -   \sum_{t=1}^{N_R-n_1}\frac{_1F_1(t;N_R;aN_RsE_D/(-bsE_D)}{(bsE_D/E_I)^t} \Gamma(t+n_2-N_R+i)\Bigg]$
\else
\begin{align*}
A(i)=\frac{1}{(n_2-n_1+i)!}\Bigg[\sum_{l=0}^{T_1}\frac{(aN_RsE_D/E_I)^l}{(N_R)_l}
\bigg[-\Gamma(n_2-N_R+i+1)\\ U(l+1,l-n_2+N_R-i+1,-bsE_D/E_I)\bigg]
	    - \;  \sum_{t=1}^{N_R-n_1}\frac{_1F_1(t;N_R;aN_RsE_D/(-bsE_D)}{(bsE_D/E_I)^t}\\
\quad \; \times \bigg[ \Gamma(t+n_2-N_R+i)\bigg]\Bigg]
\end{align*}
\fi
From \cite{detfact}, shifted factorials are defined by,
\begin{align}
(z)_{s;n}=\left\lbrace \begin{array}{ll}
1, & n=0,\\
z(z+s)....(z+(n-1)s), & n=1,2,...
\end{array}\right.
\end{align}
A special case of this is the Pochhammer's symbols, when $s=1$.
 \begin{align}
(z)_n=(z)_{1;n}=\left\lbrace \begin{array}{ll}
1, & n=0,\\
z(z+1)....(z+(n-1)), & n=1,2,...
\end{array}\right.
\end{align}
In our case, in the $\vec {\tilde N}$ matrix, we have such Pochhammer's symbols in all columns except in the first.
From \cite[Lemma.1]{detfact}, we have the relation that determinant of a matrix with $ij^{th}$ element for $ 0 \leq i,j \leq n-1$, being a shifted factorial $(z_j)_{s;i}$ is given by $|(z_j)_{s;i}|=\Delta_n(\vec z)$, where $\Delta_n(\vec z)=\prod_{0 \leq i < j \leq n-1}(z_j-z_i)$.
Evaluating $|\vec {\tilde N}|$ by Laplace expansion along the first column and using the above relation from \cite{detfact}, we get
\begin{equation}
|\vec {\tilde N}|=\sum_{i=1}^{n_1}(-1)^{i+1}A(i)\Delta_{n_1-1}^i(\vec z),
\end{equation}
where $\vec z=[n_2-n_1+1+1,n_2-n_1+1+2,....,n_2-n_1+1+n_1]$ and $\Delta_{n_1-1}^i(z)$ is the Vandermonde determinant formed by all elements of the vector $\vec z$ except the $i^{th}$ element. 
Any Vandermonde determinant remains unchanged if from each element of the matrix, one subtracts the same constant, i.e., $\Delta_n(\vec z+c)=\prod_{0 \leq k < j \leq n-1}((z_j+c)-(z_k+c))=\prod_{0 \leq k < j \leq n-1}(z_j-z_k)$. Hence, the constant $n_2-n_1+1$ can be subtracted from each element of the vector $\vec z$. Hence, 
\begin{equation}\label{Ntilde1}
|\vec {\tilde N}|=\sum_{i=1}^{n_1}(-1)^{i+1}A(i)\Delta_{n_1-1}^i(\vec z),
\end{equation}
where  $\vec z=[1,..., n_1]$.
The Vandermonde determinant $\Delta_n(\vec z)$, whose nodes are given by first $n_1$ integers, i.e., $\vec z=[1,..., n_1]$, is given by $\Delta_n(\vec z)=\prod_{1 \leq k < j \leq n_1}(j-k)$.  For simplifying this expression, we expand the double product as follows:
\begin{align}
\Delta_n(\vec z)=(n_1-1)!(n_1-2)!...(1)!=\prod_{j=1}^{n_1}(j-1)!.\label{deltaz}
\end{align}
However, we actually want to evaluate $\Delta_{n_1-1}^i(\vec z)$ and not $\Delta_n(\vec z)$. Note that  the Vandermonde determinant $\Delta_{n_1-1}^i(\vec z)$ in which the $i^{th}$ element is missing, is given by, 
$\Delta_{n_1-1}^i(\vec z)=\prod_{1 \leq k  < j \leq n_1; k,j \neq i}(j-k)$.
Note that the above expression is difficult to evaluate. Hence, to obtain a simplified expression we multiply and divide the expression for $\Delta_{n_1-1}^i(\vec z)$  by the terms that are present in $\Delta_{n_1}(\vec z)$, but are missing in $\Delta_{n_1-1}^i(\vec z)$. We thus obtain,
\begin{align*}
\Delta_{n_1-1}^i(\vec z)&=\frac{\prod_{1 \leq k  < j \leq n_1;}(j-k)}{(i-1)!(n_1-i)!.}
\end{align*}
Substituting (\ref{deltaz}) in the above expression, we obtain, $ \Delta_{n_1-1}^i(\vec z)$ in  terms of $\Delta_{n_1}(\vec z)$ as,
\begin{align*}
\Delta_{n_1-1}^i(\vec z)=\frac{\Delta_{n_1}(\vec z)}{(i-1)!(n_1-i)!}=\frac{\prod_{j=1}^{n_1}(j-1)!}{(i-1)!(n_1-i)!}.
\end{align*}
Hence the final expression becomes (\ref{laplaceint}).
\section{Moments of SINR}\label{moments}
For the case of $L=n_1$ we will derive the $l^{th}$ moment. 
The mgf equation for this case can be written as
\begin{equation}
 M_{\eta}(s)=c\sum_{k=1}^{n_1}(-1)^{k+1}\rho_k|Y_k|,
\end{equation}
where $\rho_k=\int_{0}^{\infty}(\frac{\sigma^2}{E_I}+x)e^{-x} x^{n_2-N_R}\,_0F_1(n_2-n_1+1;w_kx)
\frac{_1F_1(1;N_R;\frac{aN_Rs}{xE_I/E_D+\sigma^2/E_D-bs})}{(bsE_D/E_I-\sigma^2/E_I-x)}dx\\ \;
	    - \;  \sum_{t=1}^{N_R-n_1}\frac{_1F_1(t;N_R;aN_RsE_D/(\sigma^2-bsE_D)}{(bsE_D/E_I-\sigma^2/E_I)^t}
\; \times \bigg[\frac{\sigma^2}{E_I}\Gamma(t+n_2-N_R)
 \,_1F_1(t+n_2-N_R;n_2-n_1+1;w_k) \, \\
\quad \; + \, \Gamma(t+n_2-N_R+1)
_1F_1(t+n_2-N_R+1;n_2-n_1+1;w_i)\bigg]$, $Y_{i,j}=  \frac{\sigma^2}{E_I}\,_1F_1(n_2-j+1;n_2-n_1+1;w_k) \Gamma(n_2-j+1)+  \Gamma(n_2-j+2)
_1F_1(n_2-j+2;n_2-n_1+1;w_i)$ and $Y_k$ is the matrix $Y$ with $k^{th}$ row and first column removed.
The $l^{th}$ moment is given by $\mu_l=\frac{d^l}{ds^l}M_{\eta}(s)|_{s=0}$.
We need to evaluate $\frac{d^l}{ds^l} \rho_k$. We use the relations in \cite{chiric} to evaluate the differential and obtain,
$\frac{d^l}{ds^l} \rho_k|_{s=0}=-\alpha_{l}^{Ric}d_l$ where $\alpha_l^{Ric}= b^l \sum_{k=0}^{l}{ l \choose k} \frac{(aN_R/b)^k}{(N_R)_k}$ and
\begin{align}\label{dl}
\nonumber
d_l=&l! (\frac{E_D}{E_I})^l \sum_{n=0}^{\infty}\frac{(w_k)^n \Gamma(n_2-N_R+n+1)}{(q)_n n!}\\
\nonumber
&\quad U(l,l-n_2+N_R-n,\frac{\sigma^2}{E_I}) \\
\nonumber
& \quad + \sum_{t=1}^{N_R-n_1}(t)_l(-\frac{E_I}{\sigma^2})^t (\frac{E_D}{\sigma^2})^l\bigg[\frac{\sigma^2}{E_I}\Gamma(t+n_2-N_R)\\
\nonumber
 &\:_1F_1(t+n_2-N_R;q;w_k)\\
 &\quad  +  \Gamma(t+n_2-N_R+1) _1F_1(t+n_2-N_R+1;q;w_i)\bigg]. 
\end{align}
\section{Correlated and/or Unequal power Interferers}\label{unequal}
The determinant evaluation of $|\vec N|$ can be significantly simplified for $\sigma^2 \approx 0$. We first substitute $\sigma^2=0$ in $N_{i,j}$, to obtain
\begin{equation*}
 \mathbf N_{i,j}=\begin{cases}
	    B(i), \quad  j=1,\, i=1,..., N_I,\\
 	    r_i^{N_R-j+2}\Gamma(N_R-j+2), \quad   j=2,...,N_I, \, i=1,..., N_I.\\
         \end{cases}
\end{equation*}
where $B(i)= \sum_{l=0}^{\infty}\frac{(aN_rs)^l}{(N_R)_l}(|bsE_D|^{-l+1}U(2, 2-l, r_i|bsE_D|)) -\sum_{t=1}^{N_R-N_I}\frac{_1F_1(t;N_R;\frac{aN_Rs}{-bs})}{(bsE_D)^t} ( r_i^{t+1}\Gamma(t+1))$.  By taking $r_i^{N_r-N_I+2}$ and common gamma terms outside the determinant term we obtain,
$ M_{\eta}(s)\approx c|\mathbf{N}|$,
where
$ c=\frac{(-1)^{N_R}(\sigma^2)^{(N_R-N_I)}(-1)^{\frac{1}{2}N_I(N_I-1)}|\vec \Psi |^{-N_R}}{\prod_{i<j}^{N_I}(\frac{1}{r_i}-\frac{1}{r_j}) \prod_{k=1}^{N_I}(N_R-k)!}\\
  \qquad \times \prod_{i=1}^{N_I}r_i^{N_R-N_I+2}\prod_{j=2}^{N_I}\Gamma(N_R-j+2)$,
\begin{equation*}
 \mathbf N_{i,j}=\begin{cases}
	    B(i) r_i^{-N_R+N_I-2},\quad  j=1,\, i=1,..., N_I,\\
 	    r_i^{N_I-j}, \qquad \quad   j=2,...,N_I, \, i=1,..., N_I.\\
         \end{cases}
\end{equation*}
Expanding along the first column, we obtain an approximation for the Laplace transform for $\sigma^2=0$ as,
$ M_{\eta}(s) \approx c \sum_{i=1}^{N_I}(-1)^{i+1}B(i)r_i^{-N_R+N_I-2}|V^{i}(\vec r)|$,
where $V^{i}(\vec r)$ denotes the Vandermonde matrix formed from all elements of $\vec r=(r_1, r_2,..,r_{N_I})$ except the $i^{th}$ element. Note that, we do not substitute $\sigma^2 \approx 0$ in the $c$ term but only in the $|\vec N|$ term, to obtain the approximation.
\bibliography{bibfile}
\bibliographystyle{IEEEtran}
\end{document}